\documentclass[%
 reprint,
unsortedaddress,
 amsmath,amssymb,
aip,
 reprint,%
]{revtex4-1}

\usepackage{graphicx}% Include figure files
\usepackage{dcolumn}% Align table columns on decimal point
\usepackage{bm}
\usepackage{xcolor}
\usepackage{longtable}
\usepackage{amsmath}
\usepackage{tablefootnote}
\usepackage{siunitx}

\usepackage{dcolumn}% Align table columns on decimal point
\usepackage[utf8]{inputenc}
\usepackage[T1]{fontenc}
\usepackage{mathptmx}
\usepackage{etoolbox}

\makeatletter
\def\@email#1#2{%
 \endgroup
 \patchcmd{\titleblock@produce}
  {\frontmatter@RRAPformat}
  {\frontmatter@RRAPformat{\produce@RRAP{*#1\href{mailto:#2}{#2}}}\frontmatter@RRAPformat}
  {}{}
}%
\makeatother

\newcommand{\ket}[1]{\left|#1\right\rangle}
\newcommand{\bra}[1]{\left\langle #1\right|}

\newcommand{\hypgeo}[2]{%
  {\vphantom{F}}_{#1}\kern-\scriptspace F_{#2}%
}

\begin{document}
\title{Phenomenological characterization of the isomerization transition state of carbonyl sulfide}

\author{Amine Rafik}
\affiliation{Departamento de Ciencias Integradas y Centro de Estudios Avanzados en Física, Matemáticas y Computación, Universidad de Huelva, Huelva 21071, Spain}
\affiliation{Laboratory of Spectroscopy, Molecular Modeling, Materials, Nanomaterials, Water and Environment, LS3MN2E/CERNE2D, Faculty of Sciences, Mohammed V University in Rabat, Morocco}

\author{Jamil Khalouf-Rivera}
\affiliation{Departamento de Ciencias Integradas y Centro de Estudios Avanzados en Física, Matemáticas y Computación, Universidad de Huelva, Huelva 21071, Spain}

\author{Francisco~Pérez-Bernal}
\affiliation{Departamento de Ciencias Integradas y Centro de Estudios Avanzados en Física, Matemáticas y Computación, Unidad Asociada GIFMAN, CSIC-UHU, Universidad de Huelva, Huelva 21071, Spain}
\altaffiliation[Also at ]{Instituto Carlos I de F\'{\i}sica Te\'orica y Computacional, Universidad de Granada, Fuentenueva s/n, 18071 Granada, Spain}

\author{Khadija Marakchi}
\affiliation{Laboratory of Spectroscopy, Molecular Modeling, Materials, Nanomaterials, Water and Environment, LS3MN2E/CERNE2D, Faculty of Sciences, Mohammed V University in Rabat, Morocco}

\author{Miguel~Carvajal}
\affiliation{Departamento de Ciencias Integradas y Centro de Estudios Avanzados en Física, Matemáticas y Computación, Unidad Asociada GIFMAN, CSIC-UHU, Universidad de Huelva, Huelva 21071, Spain}
\altaffiliation[Also at ]{Instituto Carlos I de F\'{\i}sica Te\'orica y Computacional, Universidad de Granada, Fuentenueva s/n, 18071 Granada, Spain}
\email{miguel.carvajal@dfa.uhu.es}

\date{\today}

\begin{abstract}
    Signatures of excited-state quantum phase transitions in the bending degree of freedom of triatomic systems that undergo an isomerization reaction have been recently evinced. In this work, we study the carbonyl sulfide bending motion using an effective Hamiltonian within the two-dimensional limit of the vibron model framework, which has been shown to accurately describe critical phenomena in molecular bending spectra within experimental precision. To estimate the transition state energy barrier,  we propose an improvement to a phenomenological formula proposed by \citet{Baraban2015}, introducing a new term to capture the anharmonicity change that characterizes quasilinear molecules. 
\end{abstract}

\maketitle

\section{Introduction}

Transition State (TS) theory is a cornerstone in the study of chemical reactions, in particular,  to understand the energetic and structural characteristics of the transition state and to determine reaction rates~\cite{Eyring1931,Nye2007}. TS stands for a labile state  at the top of the energy barrier, in-between the reactants and the products of a reaction process. Nevertheless, despite its relevance, TS properties are chiefly calculated using different levels of theory, due to the lack of experimental information~\cite{Polanyi1995}. 

For isomerization reactions, a phenomenological formalism to determine the TS energy, based on the appearance of a dip in the energy gap of  vibrational bending states excited along the reaction coordinate, has been recently developed~\cite{Baraban2015,mellau2016}. 
However, the application of this approach is dependent on the availability of data for highly-excited levels, approaching the TS energy, in the bending vibrational band linked to the isomerization mechanism. This is a serious hindrance that can be overcome making use of accurate theoretical bending energy predictions, computed either {\it ab initio} or  using an effective Hamiltonian. 
For instance, this approach has been successfully applied to determine the barrier height of the HCN-HNC isomerization reaction using the available experimental and {\it ab initio} bending energies predicted with an effective algebraic Hamiltonian~\cite{khalou2019}. Despite the importance of the TS, the isomerization energy barrier for many simple systems remains unknown or insufficiently investigated, as it is the case for carbonyl sulphide (OCS).

Carbonyl sulfide is a quasilinear molecule which has been spectroscopically characterized in the IR region, involving rovibrational energies  up to \SI{15000}{cm^{-1}} with vibrational angular momenta $\ell=0-7$~\cite{maki1962,fayt1986global,belafhal1995,hornberger1996,YANG1997,rbaihi1998fourier,naim1998,frech1998,TRANCHART1999,GOLEBIOWSKI2014}.
In recent years, this molecular species has gained a lot of attention due to its presence in Earth’s atmosphere as well as in astronomical sources. Carbonyl sulfide is one of the most
 abundant sulfur-containing gases in the atmosphere, it has a relatively long chemical lifetime and contributes to  the greenhouse effect with a significant warming potential~\cite{bruhl2012}. This molecular species plays an important role in the formation of stratospheric sulfur aerosols and it is a possible probe for climate change due to its correlation with the biospheric uptake of CO$_2$~\cite{bruhl2012,ma2021}.
In addition, it can be used as a tracer of biogenic activity and photosynthesis~\cite{campbell2008,DeGouw2009}. In astronomical sources, OCS has been observed in a variety of systems such as planetary atmospheres, the interstellar medium, and galaxies~\cite{lellouch1995, bezard1990,jefferts1971,mauersberger1995}.

 As regards the OCS vibrational structure, a number of theoretical studies  can be found in the literature providing results  for the rovibrational structure of OCS with spectroscopic accuracy.
Variational calculations fitted the potential energy surface to the experimental energies using, e.g., different types of generalized internal coordinates~\cite{zuniga2000}, a self-consistent field-configuration interaction optimization method~\cite{xie2001}, algebraic techniques
for the analytical determination of the matrix elements~\cite{sedivcova2009}, and a polyad-conserving local algebraic model based on anharmonic ladder operators~\cite{Suarez2025}. Empirical rovibrational levels were determined using \texttt{MARVEL} from the compilation of experimental transitions~\cite{xu2024empirical}.
{\it Ab initio} methods were also applied to reproduce the energy levels and the transition lines based on higher-level calculations of the potential surface, refined empirically in some cases, and the dipole moment surface~\cite{Dobrolyubov2024,owens2024abinitio,owens2024exomol,Huang2025}. 
Up to our knowledge, there is only one study devoted to the computational modeling of the isomerization reaction of OCS~\cite{lara2018}. In this work, the authors calculated the potential surface of the ground state simulating the isomerization reactions of the isomers of OCS, mainly at the CCSD(T)/aug-cc-pVTZ level of theory. The employed statistical methods allowed for estimating the barrier height for the isomerization of OCS. The authors found that the reaction barrier from the linear OCS to its linear COS isomer goes through a stable intermediate cyclic structure $\Delta$-OCS. Despite the large number of works that have investigated the vibrational spectrum of OCS, the available levels lie well below  halfway  the estimated isomerization energy barrier.

In the present work we make use of the vibron model, a computationally efficient phenomenologic approach for the calculation of vibrational and rovibrational molecular spectra based on Lie algebras~\cite{bookalg}, that models molecular structure through collective bosonic excitations (vibrons)~\cite{iachello1981, bookmol, frank,Oss1996}. Different algebraic approaches, grounded on the original vibron model, were developed to avoid the mathematical complexity of the full rovibrational analysis in polyatomic molecules. In particular, we make use of the two-dimensional limit of the vibron model (2DVM), introduced to model vibrational bending degrees of freedom~\cite{iachello1996,Oss1996,perez2008}. 
  Despite its apparent simplicity, the 2DVM has been proven effective for describing not only the rigidly-linear and rigidly-bent limiting cases, but also the more involved quasilinear or non-rigid molecular spectra~ \cite{iachello2003,PBernal2005,perez2008}.

In the seminal work of \citet{Dixon1964}, it was shown that the crossing of the barrier to linearity is evinced in the spectrum by  a change in the pattern of energy spacings and the appearance of what was dubbed a \textit{Dixon dip}. Numerous studies have subsequently explored the energy spectrum of nonrigid bent molecules, characterized by large amplitude vibrational degrees of freedom and capable of undergoing a bent-to-linear transition. These include, for instance, investigations into quasilinearity through the definition of a quantity to determine to what degree molecules are linear or bent~\cite{Yamada1976}, or introducing quantum monodromy to explain the dependence of energies with vibrational angular momentum and the absence of a set of vibrational quantum numbers globally valid for the entire spectrum~\cite{Child1998, Child1999}.

The existence of a barrier to linearity was also associated with an excited-state quantum phase transition (ESQPT), an extension to excited states of the well-known ground-state quantum phase transitions~\cite{Caprio2008,pavels,Cejnar2021}. Experimental results for molecular bending spectra were the first systems where ESQPT precursors were identified, for which a set of characteristic spectral signatures were explained in a unified framework within the 2DVM~\cite{perez2008, larese2011, LARESE2013310, KHALOUF2021,khalouf2022}.
Afterwards, it was found that the ESQPT observed in molecular large amplitude bending modes can be extended  to
model isomerization processes. In particular, the isomerization barrier height for the HCN-HNC system was obtained from the vibrational bending spectra computed  using the 2DVM~\cite{khalou2019}. Recently, the 2DVM has been used to measure the quantum chaoticity of a system using asymptotic values of an out-of-time-ordered correlator~\cite{Novotny2023}.

In the present work, we calculate the bending spectrum of the OCS molecule using the 2DVM, optimizing the parameters of a four-body algebraic Hamiltonian with existing data, as has been done for other molecular species~\cite{KHALOUF2021}. With the resulting energies and wave functions, we have studied the OCS bending spectrum, paying special heed to the vicinities of the  ESQPT critical energy.
To that end, quantities such as the participation ratio, the expectation value of the 2DVM number operator, the quasilinearity parameter, and the effective frequency are discussed. 
Concerning the effective frequency, we also propose a new empirical formula, based on the one presented in Ref.~\cite{Baraban2015}, which reproduces the anharmonicity trend observed in quasilinear molecules. Subsequently, making use of this formula, we estimate the barrier height associated with the isomerization of the OCS species.

\section{Theoretical framework}
\subsection{The two dimensional limit of the vibron model}
The 2DVM model, that stems from the vibron model~\citep{iachello1981}, describes single (or coupled) molecular vibrational bending modes with one (or several copies of a) $U(3)$ dynamical algebra~\cite{iachello1996}. This model has been been applied to the study of the bending  spectrum of various molecular systems~\cite{iachello2003,PBernal2005,Mariano2012,SanchezCastellanos2012,Estevez-Fregoso2018,larese2011, LARESE2013310, KHALOUF2021,khalouf2022,Marisol2022,Marisol2023,Lemus2024}. Most notably, the model was shown to be capable of capturing the signatures of ground- and excited-state quantum phase transitions associated with the large-amplitude motion that occurs in bent-to-linear transitions~\cite{perez2008,larese2011,LARESE2013310, KHALOUF2021}. 
As this model has been meticulously described before (see, e.g.,  \cite{perez2008,KHALOUF2021}), in the present work we only sketch the effective 2DVM Hamiltonian and the basis set used to characterize the bending spectrum of the OCS  molecular species.

There are two exactly-solvable limits under the 2DVM theoretical framework that correspond to rigid linear and bent configurations, respectively. The rigid linear case is associated mathematically with a truncated cylindrical harmonic oscillator (the $U(3)\supset U(2)$ dynamical symmetry) and the rigid bent case can be mapped to a 2D Morse oscillator (the $U(3)\supset SO(3)$ dynamical symmetry)~\cite{perez2008}. Both subalgebra chains contain a common $SO(2)$ subalgebra with an associated quantum label,$\ell$, that is the vibrational angular momentum in the linear case and the projection along the figure-axis of the rotational angular momentum in a rigid bent molecule.
Each dynamical symmetry  provides  a possible basis set for performing the calculations. Due to the quasilinear nature of OCS, its equilibrium structure is linear and we use in our calculations the cylindrical oscillator basis. The quantum number $N$, which labels the totally-symmetric representation of $U(3)$ that spans the Hilbert space where calculations are carried out, is related to the total number of bound states of the system.  States in the rigid linear case are denoted as $\ket{[N] ; n \, \ell}\equiv \ket{n^\ell}$.  The quantum number $n$,  that is the corresponding $U(2)$ Lie algebra irrep label, indicates the number of quanta of excitation of the 2D cylindrical oscillator and, as mentioned above, $\ell$, is the system vibrational angular momentum. The branching rules for the cylindrical oscillator basis  are given by

\begin{align}
    n =& N, N-1 , N-2 , . . . , 0  \\
    \ell =& \pm n, \pm(n -2) , . . . , \pm1 ~ \text{or 0 (n = odd or even )}~.\nonumber
\end{align}

In this framework, we start with a general Hamiltonian including up to four-body interactions introduced in~\cite{KHALOUF2021}. For OCS, after a careful selection of relevant operators, we have constructed an effective bending Hamiltonian that includes operators up to two-body interactions along with a three-body term  
\begin{equation}\label{eq-hamilt}
\hat{H}=  P_{1} \hat{n}+P_{2} \hat{n}^2 + P_{3} \hat{\ell}^2+P_{4} \hat{W}^2 + P_5 \hat{n} \hat{\ell}^2
\end{equation}
The interactions considered in this Hamiltonian  can be constructed using the number operator, $\hat{n}$, which gives the number of quanta of excitation in the rigid linear limit, the vibrational angular momentum operator, ${\hat \ell}$, and the Casimir operator of the $SO(3)$ subalgebra, $\hat{W}^2$, which couples $|n^{\ell}\rangle$ with $|(n\pm2)^{\ell}\rangle$ states. More details about the physical interpretation of the operators can be found in Refs.~\cite{perez2008,KHALOUF2021}.
Despite its simplicity, the effective Hamiltonian (\ref{eq-hamilt}) allows us to carry out calculations of the OCS molecular vibrational energy structure with uncertainties close
to spectroscopic accuracy. It is block diagonal in $\ell$, as  the vibrational angular momentum is conserved~\cite{KHALOUF2021}.

\subsection{Effective frequency and isomerization transition energy}

It should be highlighted that the 2DVM, although deceptively simple, possesses significant predictive power making it possible to compute highly excited states with good accuracy~\cite{Lemus2014,Marisol2020,KHALOUF2021,larese2011,LARESE2013310}. In fact, predicted bending term values at high energies were used to estimate the transition energy barrier between the HCN/HNC isomers~\cite{khalou2019} using a method presented previously by~\citet{Baraban2015}. Specifically, 
\citet{Baraban2015} proposed a phenomenological formula for the effective frequency, $\omega^{eff}$, as a function of the midpoint energies, $\bar{E}$, the mean value of adjacent vibrational energy levels, to determine the transition state.
\begin{equation}\label{TSformula-baraban}
   \omega^{\rm eff} \left(\bar{E}\right) = \omega_0 \left(1-\frac{\bar{E}}{E_{\rm TS}}\right)^{1/m_1}~.
\end{equation}
The midpoint energies $\bar{E}$ are obtained from the predicted bending spectrum, and the three parameters of the formula, $\omega_0$, $E_{TS}$, and $m_1$, are optimized to fit $\bar{E}$ values. The $\omega_0$ parameter is the fundamental frequency ($\omega^{eff}$ at $\bar{E}=0$), $E_{TS}$ is the transition state energy, and $m_1$ is a parameter  greater or equal than $2$  which depends on the potential shape.

Nevertheless, quasilinear molecules as OCS,  that undergo  anharmonicity changes  from positive to negative values, are not well described by Eq.~\eqref{TSformula-baraban}. To encompass such cases,  we propose the addition of an extra term to Eq.~\eqref{TSformula-baraban}
\begin{equation}
\label{modified-baraban-formula}
    \omega^\text{eff}\left(\bar{E}\right) = \omega_0  \left(1 - \frac{\bar{E}}{E_{TS}}\right)^{\frac{1}{m_1}}  \left(1 + \frac{\bar{E}}{E_{TS}}\right)^{\frac{1}{m_2}}~,
\end{equation}
\noindent  with a total of four parameters:  $\omega_{\rm 0}$, $E_{\rm TS}$, $m_1$, and $m_2$. Eq.~\eqref{TSformula-baraban} is recovered when $m_2 \to \infty$. When the mid-point energy is close to zero, the first-order term of the Taylor expansion of Eq.~\eqref{modified-baraban-formula} predicts the behavior at the origin, $\left(1-x\right)^\frac{1}{m_1} \left(1+x\right)^\frac{1}{m_2}\approx 1 + \frac{m_1-m_2}{m_1m_2} x +\mathcal{O}\left(x^2\right)$. Then, assuming a positive anharmonicity at low energies, the condition $m_1>m_2$ must be satisfied.

The zero point vibrational energy (ZPVE) of a quasilinear system can be estimated following the approach presented in~\citet{Baraban2015} and using the modified formula Eq.~\eqref{modified-baraban-formula}. Once we have a continuous function that describes the midpoint energy as a function of the effective frequency, i.e., $\overline{E}(n)=\omega^\text{eff}\left(n+\frac{f}{2}\right)$, the ZPVE can be estimated solving numerically the integral 
\begin{equation}\label{eq:action}
    \int_0^{\text{ZPVE}} \frac{d\overline{E}}{\omega^\text{eff} (\overline{E})}=\frac{f}{2}~,
\end{equation}
\noindent where $f$ is the number of vibrational degrees of freedom of the system. In particular, $f=2$ for degenerate bending modes. In App.~\ref{App:ZPVE}, we solve Eq.~\eqref{eq:action} for the effective frequency formula given in Eq.~\eqref{modified-baraban-formula}.

\section{Results and Discussion}
\label{sec-results}
As  already shown in Ref.~\cite{khalou2019}, isomerization reaction barriers  can be linked to the occurrence of an ESQPT in the bending degrees of freedom.
%, appearing in-between their linear and bent structures 
Therefore, to determine the isomerization transition state energy, we can make use of  quantities that are often used to characterize ESQPTs, such as the participation ratio (PR)~\cite{LSantos2015, Santos2016, khalou2019,KHALOUF2021,khalouf2022,jamil2022}, the expectation value of the number operator (${\hat n}$)~\cite{AlvarezBajo2010,jamil2022}, the quasilinearity parameter~\cite{KHALOUF2021,jamil2022}, or the effective frequency~\cite{Baraban2015,mellau2016,KHALOUF2021,jamil2022}. Thus,  having the aim of determining the isomerization reaction barrier of OCS,  we proceed to calculate the pure bending energies and wavefunctions.

 To our knowledge, the available experimental dataset for bending vibrational levels of the main OCS isotopologue only includes states with vibrational angular momentum $\ell = 0, \ldots, 7$ ~\cite{maki1962,fayt1986global,belafhal1995,hornberger1996,YANG1997,rbaihi1998fourier,naim1998,frech1998,GOLEBIOWSKI2014}. From this comprehensive data set, 71 term values are pure bending states. Out of these, 51 of them are experimental, with energies up to \SI{8000}{cm^{-1}}~\cite{fayt1986global,belafhal1995,GOLEBIOWSKI2014}, while the remaining 20 term values are  predictions from a global analysis making use of an effective Hamiltonian, with energies up to \SI{11000}{cm^{-1}}~\cite{GOLEBIOWSKI2014}.
The characterization of the bending spectrum is carried out in this work with the effective 2DVM Hamiltonian in Eq.~\eqref{eq-hamilt}, fitting its  free parameters to the available data. 
We fit the $P_{i}$ free parameters considering different values of $N$. We manually select the value of $N$ that minimizes the deviation between calculated and both experimental and predicted data sets.
A Python code was developed to calculate the energy levels and the free parameters are optimized using the  nonlinear least squares minimization procedure provided by \texttt{lmfit} Python library~\cite{newville2016lmfit}. The code is available upon request from the authors. The quality of the fit is  assessed with the {\it rms} 
\begin{equation}
    rms = \sqrt{\frac{\sum\limits_{k=1}^{N_{\rm data}}\left(E_k^{\rm calc}-E_k^{\rm exp}\right)^2}{N_{\rm data}-n_{\rm p}}}
\end{equation}
where $E^{\rm calc}$ are the calculated energies, $E^{\rm exp}$ are the experimental energies, $N_{data}$ is the number of experimental energies, and $n_p$ is the number of free parameters considered in the fitting procedure.  In the OCS case, the 2DVM Hamiltonian $\hat{H}$ has five free parameters associated with all one- and two-body interactions plus a particular 3-body interaction. 

We carried out two fits, dubbed \textit{Fit I} and \textit{Fit II}, with a vibron number $N = 172$. In  \textit{Fit I}, the parameters are optimized taking into consideration exclusively the 51 experimental bending term values. The obtained parameter values and their uncertainties are provided in the first column of Tab.~\ref{tab:fit_results}. In this case the $rms$ is labeled as $rms(fit) = \SI{0.29}{cm^{-1}}$. If we include in the calculation the additional 20 term values obtained with an effective Hamiltonian, without further optimization, the result is $rms(pred) = \SI{1.55}{cm^{-1}}$, making clear that the predicted term values are in a reasonable agreement with the computed energy values. In fact, \citet{YANG1997} already showed that an effective Hamiltonian applied for characterizing the low vibrational levels of OCS can predict higher excited states within experimental accuracy. 
\textit{Fit II} calculation includes in the fit both experimental and predicted term values, expanding the energy range of the fit to $\SI{11000}{cm^{-1}}$. The obtained results are given in the second column of Tab.~\ref{tab:fit_results}, with an $rms(fit) = \SI{0.39}{cm^{-1}}$, which lies  very close to the expected experimental uncertainty. In this case, $rms(pred) = \SI{0.43}{cm^{-1}}$ is the $rms$ obtained considering only the experimental term values and fixing the parameters to the optimized values of \textit{Fit II}, obtaining a fit quality that is close enough to the results obtained in  \textit{Fit I}. The energies obtained in both fits, as well as the level assignment using rigid linear quantum numbers,  and the residuals with experimental and predicted (marked with an asterisk) levels is provided in Tab.~\ref{tab:computed_energies}. Calculated energy levels up to \SI{16000}{cm^{-1}} for the two reported fits are provided as supplementary data.

\begin{figure}
    \centering
    \includegraphics[width=1\linewidth]{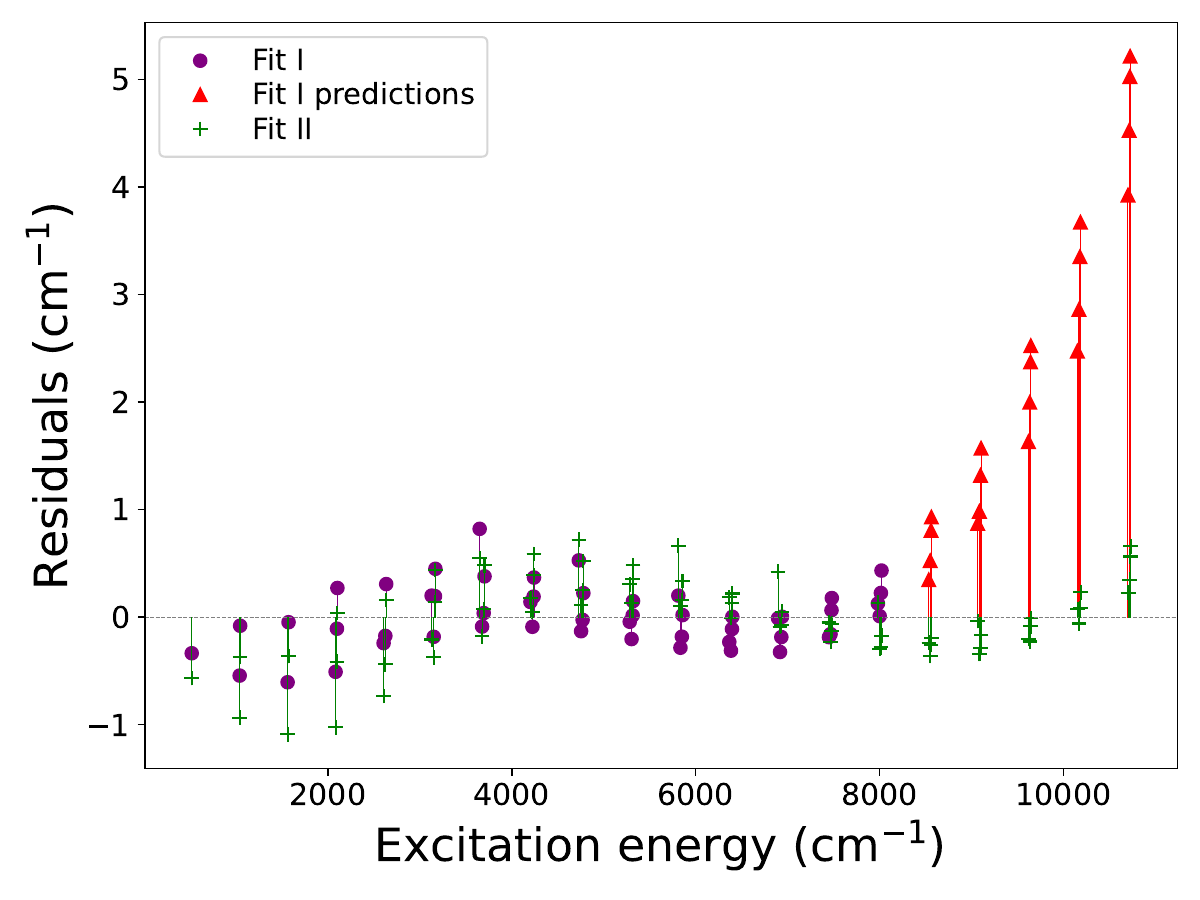}
    \caption{Residuals for \textit{Fit I} and \textit{Fit II} calculations versus the 
experimental and predicted energies in cm$^{-1}$ units. Purple circles show the residuals obtained from \textit{Fit I} for 51 experimental energies. Red triangles include the predictions of \textit{Fit I} for the 20 term values  in the energy range from \SI{8000}{cm^{-1}} to \SI{11000}{cm^{-1}} obtained with an effective Hamiltonian~\cite{GOLEBIOWSKI2014}. 
    Green crosses correspond to the residuals for \textit{Fit II}, including 71 experimental and predicted bending term values.}
    \label{fig:relative_residuals}
\end{figure}

In order to compare the results obtained from \textit{Fit I} and \textit{Fit II}, we depict the residuals, i.e., the differences between the experimental and computed energies, in cm$^{-1}$ units in Fig.~\ref{fig:relative_residuals}. \textit{Fit I} residuals for the  51 experimental term values are depicted using purple circles and the residuals for the 20 additional levels provided in  Ref.~\cite{GOLEBIOWSKI2014} are depicted with red triangles. The relative residuals obtained with \textit{Fit I} parameters for the states in the expanded dataset that lies in the range \SIrange{8000}{11000}{cm^{-1}} are below 0.05~\%,  with a largest residual $\Delta E=E^{\rm exp}-E^{\rm cal}$ of about \SI{5}{cm^{-1}}. Finally, the relative residuals  from \textit{Fit II} are depicted with green crosses. In all cases, the accuracy is below \SI{0.12}{\percent}.

\begin{table}%[htb]
    \centering
    \caption{Optimized Hamiltonian parameters for \textit{Fit I} and \textit{Fit II}, $rms$ values (see main text), and number of term values included in the fit ($N_\text{data}$). All calculations were performed for a vibron number $N=172$. Parameters' uncertainties are given in parentheses in units of the last quoted digits. Except $N_\text{data}$, all parameters and rms values are expressed in cm$^{-1}$ units.}
    \begin{tabular}{ccc}
        \toprule
           & Fit I & Fit II \\
\hline
$P_{1}$ & 2321.900(21) & 2190.543(14) \\
$P_{2}$ & -12.41738(12) & -11.58027(10) \\
$P_{3}$ & 4.3344(83) & 3.9966(75) \\
$P_{4}$ & -3.217143(46) & -3.015298(35) \\
$P_{5}$ & 0.00709(33) & 0.007085(27) \\
\hline
$N_\text{data}$ & 51 & 71 \\
$rms(fit)$ & 0.290  &  0.394 \\
$rms(pred)$ & 1.550 &  0.430\\
         \toprule
    \end{tabular}
    \label{tab:fit_results}
\end{table}

\begin{figure}%[h!]
    \centering
    \includegraphics[width=1\linewidth]{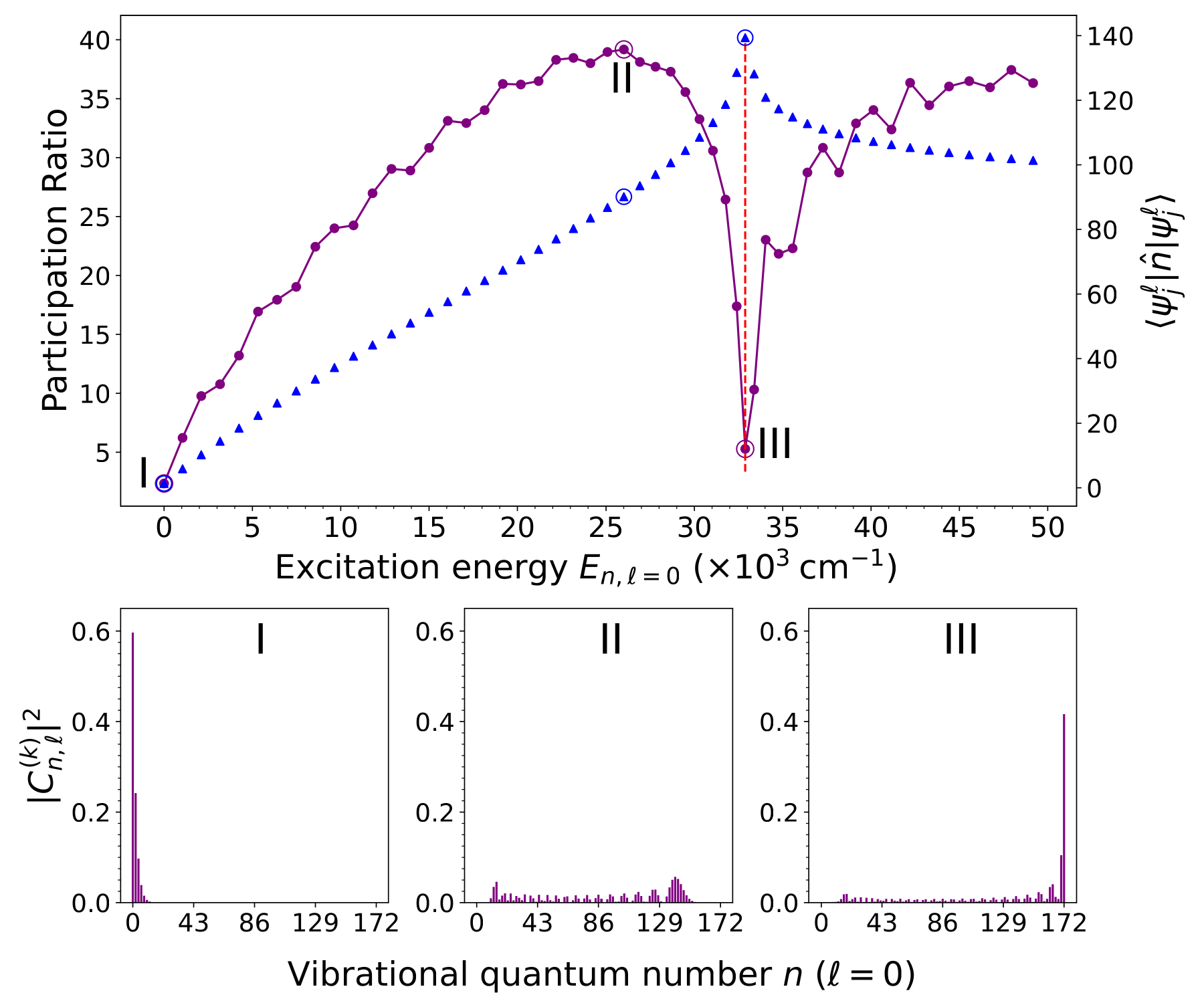}
    \caption{Upper panel: Participation ratio (purple dots) and expectation value of the $\hat{n}$ operator (blue triangles) as a function of the computed bending term values for $\ell=0$ states obtained in \textit{Fit I}. Lower panels: Squared components $|C^{(k)}_{n,\ell}|^2$  as a function of the vibrational quantum number $n$ for $\ell=0$ for three selected eigenstates (I, II, and III) marked in the upper panel with circles. }
\label{fig:pr_nExpectVal_components_H1}
\end{figure}

Following the approach of Refs.~\cite{Baraban2015, khalou2019}, we use the calculated energies and wave functions to locate the critical energy of the ESQPT, which is our estimation of the transition state energy. 
We use different quantities that are convenient probes for the ESQPT. The first one is the PR, a quantity that measures the level of localization of  wave functions in a given basis~\cite{ZELEVINSKY1996}. It was shown that the participation ratio when eigenstates are expressed in the basis associated with  a particular dynamical symmetry is a convenient probe for the location of the ESQPT critical energy~\cite{LSantos2015,Santos2016}. The participation ratio is defined as 
\begin{equation}
     PR\left(\ket{\psi_k^{\ell}}\right) = \frac{1}{\sum_{n}{|C^{(k)}_{n,\ell}|^4}}~,
\end{equation}
\noindent where $\ket{\psi_k^{\ell}}$ is an eigenstate with vibrational angular momentum $\ell$. The  eigenstates $\{\ket{\psi_k^{\ell}}\}$ can be expressed as $\ket{\psi_k^{\ell}}=\sum_{n=0} C^{(k)}_{n,\ell} \ket{n \, \ell}$. The lower the value of PR, the more localized is the state under study in the given basis set with a minimum value of unity if the eigenstate is a basis state. Conversely, the larger the value of the PR, the more spread out is the state, with a maximum value equal to the dimension of the $\ell$ angular momentum Hamiltonian block. 
A second hallmark of the ESQPT that allows us to localize the TS energy is revealed by the expectation value of the ${\hat n}$ operator, $\bra{\psi_k^{\ell}} {\hat n} \ket{\psi_k^{\ell}}$, which is the operator used as an order parameter for the second-order ground-state quantum phase transition~\cite{perez2008}.
 We depict in Fig.~\ref{fig:pr_nExpectVal_components_H1} the PR and the expectation value of ${\hat n}$ for the  $\ell = 0$ eigenfunctions obtained in \textit{Fit I} as a function of the calculated term values in cm$^{-1}$ units. 
The purple line is the PR and the blue triangles mark the expectation value of $\hat{n}$. 
The PR results in Fig.~\ref{fig:pr_nExpectVal_components_H1} indicate that, as expected, eigenstates are more delocalized in the basis as the energy increases from the ground state. This is followed by a striking decrease to a minimum PR value, observed around \SI{33000}{cm^{-1}}, that marks the ESQPT critical energy~\cite{LSantos2015,Santos2016,KHALOUF2021} which in this case is associated with the isomerization reaction transition state energy~\cite{khalou2019}.
Three eigenstates of OCS with $\ell=0$, designated as I, II, and III, have been chosen to examine the eigenstate structure at different excitation energies. In particular, eigenstate I is the ground state, eigenstate II lies at an energy halfway between the energy of states I and III, and eigenstate III is the eigenstate with a minimum PR value. The squared wave function components,  $|C^{(k)}_{n,\ell}|^2$, 
of these three eigenstates as a function of the quantum number $n$ in the $\ket{n \, \ell}$ basis  are shown as bar plots in the three lower panels of Fig.~ \ref{fig:pr_nExpectVal_components_H1}. 
The bar plots indicate that eigenstates I and III are significantly localized while eigenstate II is delocalized in the states of the basis set, confirming the PR results. Wavefunction I has a squared component around $0.6$ in the basis state $\ket{n=0 \,, \ell=0}$
and wavefunction III is localized in the last basis state, $\ket{n=N \,, \ell=0}$, of the basis set with a squared component close to $0.4$. The minimum PR value is linked to the existence of an ESQPT, which is in fact connected to  anharmonicity effects in the Hamiltonian~\cite{AlvarezBajo2010,jamil2022} and associated with the isomerization transition in the system~\cite{khalou2019}.
Regarding the expectation value of the number operator, $\hat{n}$, as a probe for the system's isomerization, this is evaluated for the different system eigenstates. This quantity is depicted in Fig.~\ref{fig:pr_nExpectVal_components_H1} as a function of the excitation energy, and its maximum value is located at the ESQPT critical energy, a result that agrees with the PR result. The red dashed vertical line indicates the isomerization TS energy obtained with the modified empirical formula proposed for the effective frequency $\omega^{\rm eff}$ in Eq.~\eqref{modified-baraban-formula}. The PR and expected values of ${\hat n}$ for the eigenfunctions of \textit{Fit II} are reported in the supplementary material.

\begin{figure}%[htb]
    \centering
    \includegraphics[width=0.8\linewidth]{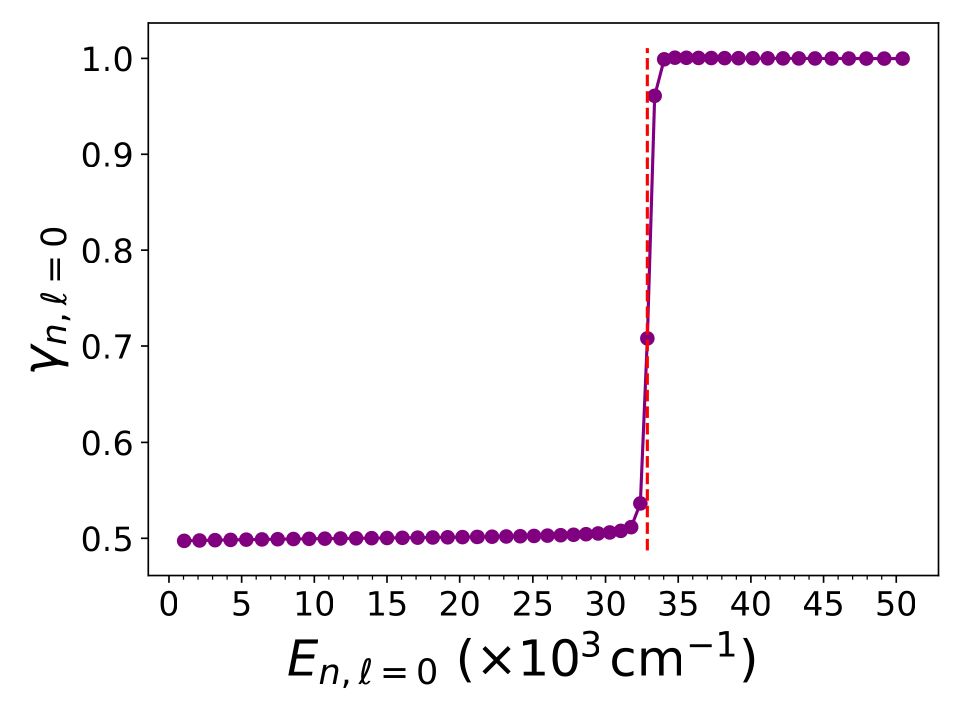}
    \caption{Quasilinearity parameter $\gamma_{n,\ell=0}$ as a function of the bending energies calculated using the Hamiltonian (\ref{eq-hamilt}) in {\it Fit I}.}
    \label{fig:quasilinearity_H1}
\end{figure}

Another quantity of interest in the case of molecular bending vibrations is the quasilinearity parameter
\begin{equation}
    \gamma_{n,\ell}=\frac{E_{n+1, \ell+1} - E_{n,\ell}}{E_{n+2, \ell} - E_{n, \ell}} ~~,
\end{equation}
\noindent originally proposed by \citet{Yamada1976} to examine the degree of quasilinearity of the molecular bending modes and recently extended to the study of ESQPTs in the bending vibration of molecules~\cite{jamil2022}.
 In this framework, the quasilinearity parameter is equal to $0.5$ for the bending mode of a rigidly-linear molecule with symmetry ${\cal C}_{\infty v}$ or ${\cal D}_{\infty h}$. In the OCS case, the  $\gamma_{n,0}$ parameter as a function of the excitation bending energies for \textit{Fit I} results is shown in Fig.~\ref{fig:quasilinearity_H1}. The value of $\gamma_{n,0} = 0.5$ for the OCS ground and lowest excited states clearly indicate that this molecular species  can be considered a linear molecule within this energy range. A sudden rise of the value of $\gamma_{n,0}$ to a value of $1$ occurs in the region of \SI{33000}{cm^{-1}}, at the ESQPT critical energy~\cite{jamil2022}. We associate this OCS structural change with the isomerization transition state energy,  a result that is consistent with the isomerization energy barrier value estimated with the PR and the expectation value of the number operator. 

\begin{figure}%[htb]
    \centering
    \includegraphics[width=1\linewidth]{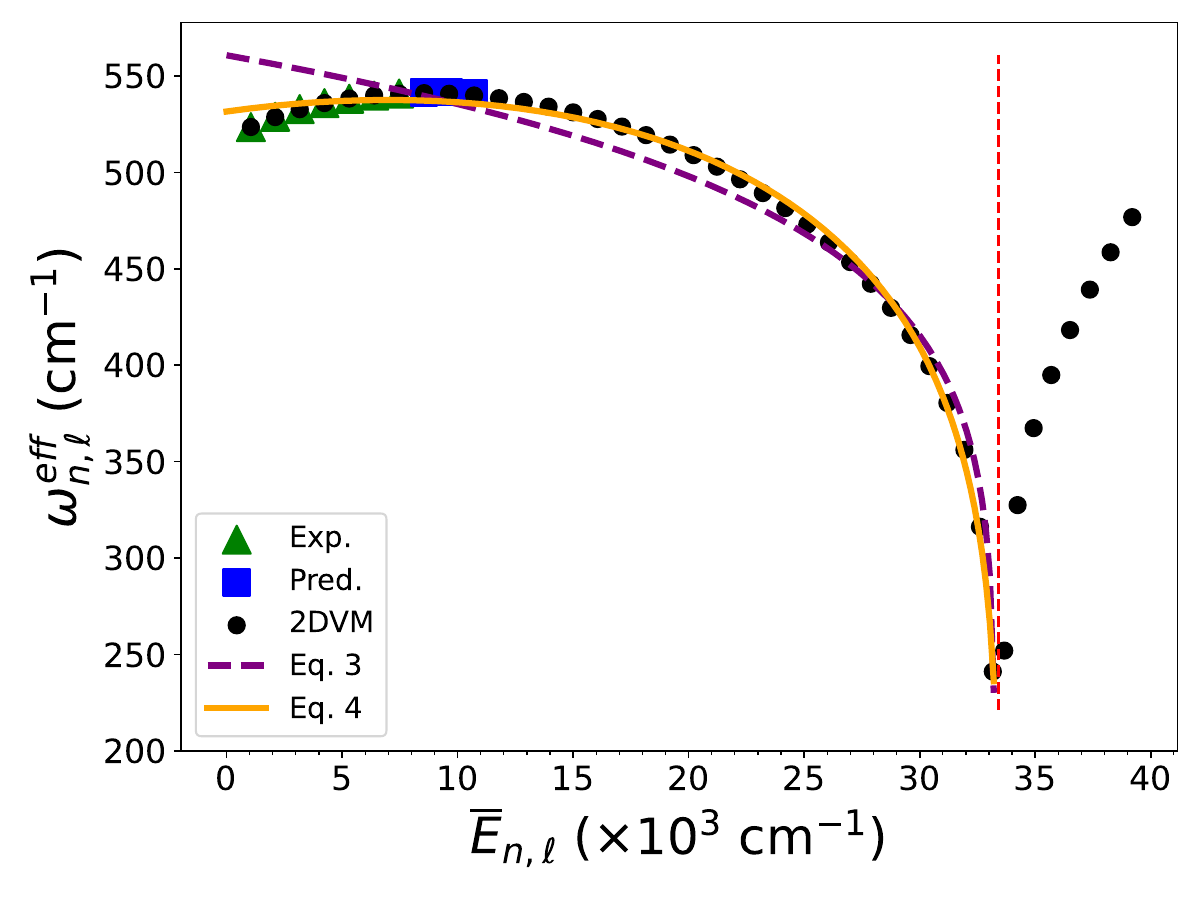}
    \caption{Effective frequency as a function of the midpoint bending energy for $\ell=0$ states computed  for \textit{Fit I} results. Green triangles and blue squares are the available experimental and extended energy data. Black circles are the 2DVM results. Purple and orange lines are obtained fitting Eqs.~\eqref{TSformula-baraban} and (\ref{modified-baraban-formula}) to \textit{Fit I} results, respectively.}  \label{fig:effec_freq_baraban_and_new_fomula}
\end{figure}

From the results obtained for the PR, the expected value of $\hat n$, and the quasilinearity parameter, we have a broad estimation of the transition state energy in the range \SI{33000}{cm^{-1}}. Nevertheless, a more precise estimate can be obtained based on the variation of the effective frequency $\omega^{\rm eff}$, computed from the 2DVM predictions for the bending energies of OCS.  In Fig.~\ref{fig:effec_freq_baraban_and_new_fomula}, we plot the values of $\omega^{eff}$ obtained from the 2DVM predictions with $\ell=0$ of \textit{Fit I} (black circles) as a function of the midpoint bending energies. We have also included the available experimental data, lying in the energy range \SIrange{0}{8000}{cm^{-1}} (green triangles), and the extended data with $\ell=0$, found in the \SIrange{8000}{11000}{cm^{-1}} energy range (blue squares).
We would like to emphasize that the values of $\omega^{eff}$ predicted from the spectrum of \textit{Fit I} plunge as the excitation energy approaches the expected isomerization barrier energy, providing an estimate of this critical parameter.

A finer estimate of the isomerization barrier of OCS can be determined using a phenomenological approach, as proposed by~\citet{Baraban2015}. This approach allows us to determine the TS energy more accurately by fitting  the $\omega_{\rm 0}$, $E_{\rm TS}$, and $m_1$ parameters  in the phenomenological formula \eqref{TSformula-baraban} to optimize the agreement with the computed effective frequency values. The purple line in Fig.~\ref{fig:effec_freq_baraban_and_new_fomula}  corresponds to the results of this fit (\ref{TSformula-baraban}), which provides a reasonable description of the effective frequency dip. 

However, improved results can be obtained from the extended equation~\eqref{modified-baraban-formula}. The fit of this formula to the midpoint energies predicted by \textit{Fit 1} is shown as an orange line in Fig.~\ref{fig:effec_freq_baraban_and_new_fomula}. Comparing the results obtained with both phenomenological equations, it is clear that, at the cost of including the new parameter $m_2$,  the results obtained with Eq.~\eqref{modified-baraban-formula} have a better agreement with effective frequency values at low energies. Hence, the new empirical formula Eq.~(\ref{modified-baraban-formula}) corrects  the trend of the effective frequency for molecules with quasilinear character. 

A summary of the results obtained in the present work for the OCS TS energy and ZPVE, as well as the  values in the literature obtained with {\it ab initio} calculations can be found  in Table~\ref{tab:computed_TS_ZVPE_energies}.
The values of $E_{\rm TS}$ and ZPVE from \textit{ab initio} calculations were calculated at a CCSD(T)/aug-cc-pVTZ level using \texttt{Gaussian G09} and \texttt{CFOUR}~\cite{lara2018}. 
As regards our results, we provide the results obtained optimizing the parameters in effective frequency formulas in Eqs.~(\ref{TSformula-baraban}) and (\ref{modified-baraban-formula}) to the 2DVM bending energy predictions from \textit{Fit I} and \textit{Fit II}. The  values $E_{\rm TS}$ from \textit{Fit I} results and Eqs.~(\ref{TSformula-baraban}) and (\ref{modified-baraban-formula}) are 33240 and 33409~cm$^{-1}$, respectively. The results obtained considering  \textit{Fit II} data are 35769 and 36005~cm$^{-1}$. 
The barrier to OCS isomerization determined by this procedure is in good agreement with the {\it ab initio} results estimated at 33052/33611 cm$^{-1}$ using G09 and CFOUR~\cite{lara2018}. 
The differences of the $E_{\rm TS}$ value obtained for OCS with \textit{Fit I} and \textit{Fit II} compared to  {\it ab initio} G09 and CFOUR calculations are 0.6\% and 1.1\%, and 8.2\% and 6.4\% respectively for Eq.~\eqref{TSformula-baraban}, and 1.1\% and 0.6\%, and 
8.9\% and 7.1\% respectively for Eq.~(\ref{modified-baraban-formula}). Therefore, it can be highlighted that the combination of the computationally inexpensive 2DVM approach, with phenomenological formulas (\ref{TSformula-baraban}) and (\ref{modified-baraban-formula}), can provide a reasonable estimation  of the isomerization barrier of OCS, in line with results calculated with the golden standard method CCSD(T). 
As regards the ZPVE estimation, the results obtained with 
Eq.~(\ref{modified-baraban-formula}),  using \textit{Fit I} and \textit{Fit II} results, are notably closer to the known estimates than the results obtained using Eq.~(\ref{TSformula-baraban}) from Ref.~\cite{Baraban2015}.  
The maximum deviation of ZPVE in both fits with respect to the {\it ab initio} bending fundamental is about 8.7\% using Eq.~\eqref{TSformula-baraban} and 2.2\% for Eq.~(\ref{modified-baraban-formula}).
In addition, the ancillary optimized parameters $\omega_0$, $m_1$ and $m_2$ are also provided in Table~\ref{tab:computed_TS_ZVPE_energies}.

 Thus, considering the TS results obtained for the OCS molecule, we can conclude that the optimization of Eq.~(\ref{modified-baraban-formula}) parameters to match the 2DVM predictions provides an accurate estimation of the isomerization transition state energy, even when the available levels lie well below half the estimated isomerization energy barrier.

\section{Conclusions}
\label{sec-conclusions}

The bending spectrum of the OCS molecule is calculated within experimental accuracy using the 2DVM approach. 
Despite the lack of experimental data for highly-excited bending levels, this algebraic model successfully predicts the spectrum with sufficient accuracy to estimate the isomerization barrier, yielding results comparable to those obtained from {\it ab initio} methods. The spectroscopic signatures of the TS in OCS have been characterized by several quantities: the participation ratio, the expectation value of the ${\hat n}$ operator, the quasilinearity parameter, and the effective frequency. The agreement between these indicators and \textit{ab initio} computed TS properties highlights the quality of the bending energies and wavefunctions obtained using the 2DVM approach.

Furthermore, the isomerization barrier height estimated using the semiempirical formula developed by \citet{Baraban2015} aligns well with \textit{ab initio} results. However, this formula does not adequately capture the tendency of quasilinear molecules to exhibit a sign change in anharmonicity—from positive to negative— as the energy approaches the linearity barrier. In the case of OCS, this change is observed at $\nu_b = (n - |\ell|)/2 = 7$, where the slope of the effective frequency reverses. To address this, we propose a new phenomenological formula for the effective frequency [Eq.~(\ref{modified-baraban-formula})], which better accounts for the anharmonic behavior at low vibrational states. This improved formula yields a more accurate fit to the effective frequencies and provides an isomerization barrier height in very good agreement with the \textit{ab initio} golden standard.

Considering the promising results obtained for the bending spectrum of OCS and its transition state energy, in next works we will model algebraically the full vibrational spectrum of carbonyl sulfide  (see, e.g., Ref.~\cite{Marisol2023,Lemus2024}) in order to delve deeper into the influence of  vibrational excited states on the isomerization barrier height.

\begin{acknowledgments}
This research has received funding from the European Union’s Horizon 2020 research and innovation program under the Marie Skodowska-Curie Grant Agreement No. 872081 and Grant No. PID2022-136228NB-C21 funded by
MICIU/AEI/10.13039/501100011033 and, as appropriate, by “ERDF A way of making Europe, by ERDF/EU,”
by the European Union, or by the European Union NextGenerationEU/PRTR. This work was also partially supported by the Consejería de  Universidad, Investigación e Innovación, Junta de Andalucía and European Regional Development Fund (ERDF 2021–2027)
under the project EPIT1462023.  Computing resources supporting this work were provided by the CEAFMC and Universidad de Huelva High Performance Computer located in the Campus Universitario “El Carmen” and funded by FEDER/MINECO Project No. UNHU-15CE-2848.
\end{acknowledgments}

%%%%%%%%%%%%%%%%%%%%%%%%%%%%%%%%%%%%%%%%%%%%%%%%%%%%%%%%%%%%%

\begin{table*}
    \caption{Experimental ($E^{\rm exp}$) and computed bending energies ($E^{\rm cal}$) obtained from \textit{Fit I} and \textit{Fit II} expressed in cm$^{-1}$ units. Residual values, $\Delta E=E^{\rm exp}-E^{\rm cal}$, are also provided.} \label{tab:computed_energies} 
    \begin{tabular}{cccccc|cccccc}
\toprule
    \textbf{($n,\ell$)} & $E^{\rm exp}$\footnote{\label{IIa}As experimental data are considered those collected from Refs.~\cite{fayt1986global,belafhal1995} and the predictions in the range \SIrange{0}{8000}{cm^{-1}} computed with an effective Hamiltonian from Ref.~\cite{GOLEBIOWSKI2014}. The predicted energies in the range \SIrange{8000}{11000}{cm^{-1}}~\cite{GOLEBIOWSKI2014}
are marked with an asterisk. 
\textit{Fit I} only considers the experimental data but compares their higher excited energy calculations with those marked with an asterisk. \textit{Fit II} uses both the experimental and predicted term values marked with an asterisk. } & $E^{\rm cal}$(\textit{Fit I}) & $\Delta E$ & $E^{\rm cal}$ (\textit{Fit II}) & $\Delta E$ & ($n,\ell$) & $E^{\rm exp}$\footref{IIa} & $E^{\rm cal}$(\textit{Fit I}) & $\Delta E$ & $E^{\rm cal}$ (\textit{Fit II}) & $\Delta E$ \\
    \hline
    (0, 0) & 0.0 & 0.0 & 0.0 & 0.0 & 0.0    & (13, 3) & 6931.30\cite{GOLEBIOWSKI2014} & 6931.48 & -0.19 & 6931.37 & -0.08 \\
(2, 0) & 1047.04~\cite{fayt1986global}& 1047.12 & -0.08 & 1047.41 & -0.37  & (15, 3) & 8014.46\cite{GOLEBIOWSKI2014} & 8014.23 & 0.22 & 8014.74 & -0.28  \\
(4, 0) & 2104.83~\cite{fayt1986global} & 2104.56 & 0.27 & 2104.79 & 0.04  & *(17, 3) & 9098.23\cite{GOLEBIOWSKI2014} & 9096.91 & 1.32 & 9098.51 & -0.29 \\
(6, 0) & 3170.64~\cite{belafhal1995} & 3170.19 & 0.45 & 3170.21 & 0.44  & *(19, 3) & 10181.71\cite{GOLEBIOWSKI2014} & 10178.36 & 3.35 & 10181.63 & 0.08 \\
(8, 0) & 4242.55~\cite{belafhal1995} & 4242.18 & 0.37 & 4241.96 & 0.59 & (4, 4) & 2084.37~\cite{belafhal1995} & 2084.88 & -0.51 & 2085.4 & -1.03\\
(10, 0) & 5319.01~\cite{GOLEBIOWSKI2014} & 5318.87 & 0.15 & 5318.53 & 0.48  & (6, 4) &  3151.68\cite{GOLEBIOWSKI2014} & 3151.86 & -0.18 & 3152.05 & -0.38 \\
(12, 0) & 6398.75~\cite{GOLEBIOWSKI2014} & 6398.75 & 0.0 & 6398.54 & 0.22  &(8, 4) & 4225.02~\cite{belafhal1995} & 4225.11 & -0.09 & 4224.97 & 0.05 \\
(14, 0) & 7480.64~\cite{GOLEBIOWSKI2014} & 7480.47 & 0.18 & 7480.71 & -0.06  &(10, 4) & 5302.79\cite{GOLEBIOWSKI2014} & 5302.99 & -0.21 & 5302.65 & 0.13  \\
*(16, 0) & 8563.67\cite{GOLEBIOWSKI2014} & 8562.74 & 0.93 & 8563.86 & -0.2 &(12, 4) & 6383.71\cite{GOLEBIOWSKI2014} & 6384.02 & -0.31 & 6383.72 & -0.01  \\
*(18, 0) & 9646.90\cite{GOLEBIOWSKI2014} & 9644.38 & 2.52 & 9646.91 & -0.01 &(14, 4) & 7466.67\cite{GOLEBIOWSKI2014} & 7466.83 & -0.16 & 7466.9 & -0.23 \\
*(20, 0) & 10729.45\cite{GOLEBIOWSKI2014} & 10724.24 & 5.21 & 10728.8 & 0.66 &*(16, 4) & 8550.67\cite{GOLEBIOWSKI2014} & 8550.15 & 0.52 & 8551.03 & -0.36   \\
(1, 1) & 520.42 ~\cite{fayt1986global}&  520.76 & -0.34 & 520.99 & -0.57 &*(18, 4) & 9634.79\cite{GOLEBIOWSKI2014} & 9632.8 & 2.0 & 9635.02 & -0.23  \\
(3, 1) & 1573.37~\cite{fayt1986global} & 1573.41 & -0.05 & 1573.73 & -0.36 &*(20, 4) & 10718.17\cite{GOLEBIOWSKI2014} & 10713.64 & 4.52 & 10717.82 & 0.35 \\
(5, 1) & 2635.59~\cite{fayt1986global} &  2635.28 & 0.31 & 2635.43 & 0.16 &(5, 5) & 2606.57~\cite{GOLEBIOWSKI2014} & 2606.82 & -0.24 & 2607.3 & -0.73 \\
(7, 1) & 3704.77~\cite{belafhal1995} &  3704.39 & 0.38 & 3704.29 & 0.48 &(7, 5) &3677.79\cite{GOLEBIOWSKI2014} &  3677.88 & -0.09 & 3677.97 & -0.18  \\
(9, 1) & 4779.23~\cite{GOLEBIOWSKI2014} & 4779.0 & 0.22 & 4778.7 & 0.52 &(9, 5) & 4754.19\cite{GOLEBIOWSKI2014} &  4754.33 & -0.13 & 4754.08 & 0.11  \\
(11, 1) & 5857.56~\cite{GOLEBIOWSKI2014} &  5857.54 & 0.02 & 5857.22 & 0.33  &(11, 5) & 5834.33\cite{GOLEBIOWSKI2014} &  5834.61 & -0.29 & 5834.23 & 0.1  \\
(13, 1) & 6938.58~\cite{GOLEBIOWSKI2014} & 6938.58 & 0.0 & 6938.53 & 0.05  &(13, 5) & 6916.99\cite{GOLEBIOWSKI2014} &  6917.32 & -0.33 & 6917.08 & -0.09 \\
(15, 1) & 8021.22~\cite{GOLEBIOWSKI2014} & 8020.79 & 0.43 & 8021.4 & -0.17 &     (15, 5) & 8001.14\cite{GOLEBIOWSKI2014} &  8001.14 & 0.01 & 8001.44 & -0.29   \\
*(17, 1) &  9104.52\cite{GOLEBIOWSKI2014} &  9102.95 & 1.57 & 9104.69 & -0.17   &*(17, 5) & 9085.83\cite{GOLEBIOWSKI2014} & 9084.85 & 0.98 & 9086.17 & -0.34  \\
*(19, 1) &  10187.57\cite{GOLEBIOWSKI2014} & 10183.9 & 3.67 & 10187.34 & 0.23 &*(19, 5) & 10170.15\cite{GOLEBIOWSKI2014} & 10167.29 & 2.86 & 10170.21 & -0.06  \\
(2, 2) & 1041.29~\cite{fayt1986global}&  1041.84 & -0.54 & 1042.23 & -0.94 &(6, 6) & 3129.21\cite{GOLEBIOWSKI2014} &  3129.01 & 0.2 & 3129.42 & -0.21 \\   
(4, 2) & 2099.52~\cite{fayt1986global} & 2099.63 & -0.11 & 2099.94 & -0.41 &(8, 6) & 4203.97\cite{GOLEBIOWSKI2014} &  4203.83 & 0.14 & 4203.8 & 0.18  \\   
(6, 2) & 3165.80~\cite{fayt1986global} & 3165.61 & 0.19 & 3165.66 & 0.14 &(10, 6) & 5283.16\cite{GOLEBIOWSKI2014} & 5283.2 & -0.04 & 5282.86 & 0.3 \\  
(8, 2) & 4238.10~\cite{belafhal1995} & 4237.91 & 0.19 & 4237.71 & 0.39  &(12, 6) & 6365.42\cite{GOLEBIOWSKI2014} & 6365.65 & -0.23 & 6365.23 & 0.18  \\
(10, 2) & 5314.91~\cite{GOLEBIOWSKI2014} & 5314.89 & 0.01 & 5314.56 & 0.35  &(14, 6) & 7449.63\cite{GOLEBIOWSKI2014} & 7449.82 & -0.19 & 7449.68 & -0.05  \\        
(12, 2) & 6394.95~\cite{GOLEBIOWSKI2014} & 6395.07 & -0.11 & 6394.83 & 0.13  &*(16, 6) & 8534.79\cite{GOLEBIOWSKI2014} & 8534.45 & 0.35 & 8535.03 & -0.24  \\
(14, 2) & 7477.12~\cite{GOLEBIOWSKI2014} & 7477.06 & 0.06 & 7477.25 & -0.13  &*(18, 6) & 9619.99\cite{GOLEBIOWSKI2014} & 9618.36 & 1.63 & 9620.19 & -0.2  \\ 
*(16, 2) & 8560.39\cite{GOLEBIOWSKI2014} &  8559.59 & 0.8 & 8560.65 & -0.26 &*(20, 6) & 10704.36\cite{GOLEBIOWSKI2014} & 10700.44 & 3.92 & 10704.13 & 0.23 \\ 
*(18, 2) & 9643.85\cite{GOLEBIOWSKI2014} & 9641.48 & 2.37 & 9643.93 & -0.09 &(7, 7) & 3652.28\cite{GOLEBIOWSKI2014} &  3651.46 & 0.82 & 3651.73 & 0.55  \\   
*(20, 2) & 10726.61\cite{GOLEBIOWSKI2014} & 10721.59 & 5.02 & 10726.05 & 0.56  &(9, 7) & 4730.27\cite{GOLEBIOWSKI2014} &  4729.74 & 0.53 & 4729.55 & 0.72 \\
(3, 3) & 1562.61~\cite{fayt1986global} & 1563.22 & -0.61 & 1563.7 & -1.09 &(11, 7) & 5811.96\cite{GOLEBIOWSKI2014} & 5811.76 & 0.2 & 5811.3 & 0.66  \\  
(5, 3) & 2625.61~\cite{fayt1986global} & 2625.78 & -0.18 & 2626.04 & -0.44 &(13, 7) & 6896.12\cite{GOLEBIOWSKI2014} & 6896.13 & -0.01 & 6895.7 & 0.42 \\
(7, 3) & 3695.58~\cite{belafhal1995} & 3695.54 & 0.04 & 3695.51 & 0.07 &(15, 7) & 7981.67\cite{GOLEBIOWSKI2014} & 7981.55 & 0.13 & 7981.54 & 0.13 \\ 
(9, 3) & 4770.74~\cite{GOLEBIOWSKI2014} & 4770.77 & -0.03 & 4770.49 & 0.25 &*(17, 7) & 9067.67\cite{GOLEBIOWSKI2014} & 9066.8 & 0.87 & 9067.7 & -0.04  \\
(11, 3) & 5849.71\cite{GOLEBIOWSKI2014} & 5849.89 & -0.18 & 5849.55 & 0.15  &*(19, 7) & 10153.21\cite{GOLEBIOWSKI2014} & 10150.74 & 2.47 & 10153.13 & 0.08\\ 
        \toprule
    \end{tabular}
\end{table*}

\begin{table*}%[h]%[htb]
    \centering
    \caption{Quantitative comparison of {\it ab initio} TS and ZPVE (in units of cm$^{-1}$) and our results computed with the effective frequency formula  from Ref.~\cite{Baraban2015} and Eq.~(\ref{modified-baraban-formula}) using the 2DVM bending energy predictions from Fit I and Fit II. Uncertainties are given in parentheses in units of the last quoted digits.  }
    \begin{tabular}{ccc|ccc}
        \toprule
            Source &  TS   & ZPVE  & $\omega_0$ & $m_1$ & $m_2$\\
            \hline
            G09/CFOUR\footnote{\label{iiib}{\it Ab initio} results calculated at CCSD(T)/aug-cc-pVTZ level of theory using Gaussian G09 and CFOUR~\cite{lara2018}. ZPVE for the bending spectrum is estimated as the fundamental energy $\omega_{\rm b}$.}    & 33052/33611 & 520.6/524.5  & --- & --- & ---\\
            Fit I, Eq.~(\ref{TSformula-baraban})\footnote{The phenomenological formula  from Ref.~\cite{Baraban2015} given in Eq.\eqref{TSformula-baraban}}  &  33240(24) & 560.5(38) & 560.8(38) & 7.76(38)  & ---    \\
            Fit I, Eq.~(\ref{modified-baraban-formula})  & 33409(37)  & 532.1(21) & 531.6(21) & 4.99(15) & 3.28(22) \\
            Fit II, Eq.~(\ref{TSformula-baraban})\footref{iiib} & 35769(44)  & 565.8(48)  & 566.2(49) & 6.78(38) & ---     \\
            Fit II, Eq.~(\ref{modified-baraban-formula}) & 36005(65)  & 528.7(29) & 528.1(30) & 4.16(16) & 2.52(19)  \\   
	\toprule
    \end{tabular}
\label{tab:computed_TS_ZVPE_energies}
\end{table*}

\appendix
\section{\label{App:ZPVE} Zero point vibrational energy}

To estimate the ZPVE we have followed the methodology presented by~\citet{Baraban2015}. The integral in Eq.~\eqref{eq:action}, considering the effective frequency given by Eq.~\eqref{TSformula-baraban}, is analytically solvable and, therefore, we can find an expression for the ZPVE (see the supplementary material of Ref.~\cite{Baraban2015} for further information). However, we cannot provide an explicit formula when we consider the modified parametrization given by Eq.~\eqref{modified-baraban-formula}. In this case, the resulting integral is
\begin{widetext}
    \begin{align}
    \int_0^{\text{ZPVE}} \frac{d\overline{E}}{\omega^\text{eff} (\overline{E})}=&\frac{1}{\omega_0}\int_0^{\text{ZPVE}} \left(1-\frac{\overline{E}}{E_\text{TS}}\right)^{\frac{-1}{m_1}} \left(1+\frac{\overline{E}}{E_\text{TS}}\right)^{\frac{-1}{m_2}}d\overline{E} \nonumber\\
    =&\frac{E_\text{TS}}{2^{\frac{1}{m_1}}\omega_0} \frac{m_2}{\left(m_2-1 \right)} \left[-\hypgeo{2}{1}\left(\frac{1}{m_1},\frac{m_2-1}{m_2},\frac{2m_2-1}{m_2},\frac{1}{2}\right) \right. \label{eq:zpve}\\ 
    & \left.+\left(\frac{\text{ZPVE}+E_\text{TS}}{E_\text{TS}}\right)^{\frac{m_2-1}{m_2}} \hypgeo{2}{1}\left(\frac{1}{m_1},\frac{m_2-1}{m_2},\frac{2m_2-1}{m_2},\frac{\text{ZPVE}+E_\text{TS}}{2E_\text{TS}}\right)\right] \nonumber\\
    =&\frac{f}{2}~,\nonumber
\end{align}
\end{widetext}
where $\hypgeo{2}{1}\left(a,b,c,d\right)$ is the hypergeometric function and $f=2$ in the case of the degenerate bending modes.

The values of the unknown parameters are determined by an iterative minimization process. First, we estimate $\omega_0$, $E_\text{TS}$, $m_1$, and $m_2$ in Eq.~(\ref{modified-baraban-formula}) without a zero point energy. Then, we solve numerically Eq.~\eqref{eq:zpve} to determine a first approach to the ZPVE and, using this value, we shift the energies and repeat this process recursively until we reach the desired tolerance.

\bibliography{OCSbib}% Produces the bibliography via BibTeX.

\begin{thebibliography}{69}%
\makeatletter
\providecommand \@ifxundefined [1]{%
 \@ifx{#1\undefined}
}%
\providecommand \@ifnum [1]{%
 \ifnum #1\expandafter \@firstoftwo
 \else \expandafter \@secondoftwo
 \fi
}%
\providecommand \@ifx [1]{%
 \ifx #1\expandafter \@firstoftwo
 \else \expandafter \@secondoftwo
 \fi
}%
\providecommand \natexlab [1]{#1}%
\providecommand \enquote  [1]{``#1''}%
\providecommand \bibnamefont  [1]{#1}%
\providecommand \bibfnamefont [1]{#1}%
\providecommand \citenamefont [1]{#1}%
\providecommand \href@noop [0]{\@secondoftwo}%
\providecommand \href [0]{\begingroup \@sanitize@url \@href}%
\providecommand \@href[1]{\@@startlink{#1}\@@href}%
\providecommand \@@href[1]{\endgroup#1\@@endlink}%
\providecommand \@sanitize@url [0]{\catcode `\\12\catcode `\$12\catcode `\&12\catcode `\#12\catcode `\^12\catcode `\_12\catcode `\%12\relax}%
\providecommand \@@startlink[1]{}%
\providecommand \@@endlink[0]{}%
\providecommand \url  [0]{\begingroup\@sanitize@url \@url }%
\providecommand \@url [1]{\endgroup\@href {#1}{\urlprefix }}%
\providecommand \urlprefix  [0]{URL }%
\providecommand \Eprint [0]{\href }%
\providecommand \doibase [0]{http://dx.doi.org/}%
\providecommand \selectlanguage [0]{\@gobble}%
\providecommand \bibinfo  [0]{\@secondoftwo}%
\providecommand \bibfield  [0]{\@secondoftwo}%
\providecommand \translation [1]{[#1]}%
\providecommand \BibitemOpen [0]{}%
\providecommand \bibitemStop [0]{}%
\providecommand \bibitemNoStop [0]{.\EOS\space}%
\providecommand \EOS [0]{\spacefactor3000\relax}%
\providecommand \BibitemShut  [1]{\csname bibitem#1\endcsname}%
\let\auto@bib@innerbib\@empty
%</preamble>
\bibitem [{\citenamefont {Baraban}\ \emph {et~al.}(2015)\citenamefont {Baraban}, \citenamefont {Changala}, \citenamefont {Mellau}, \citenamefont {Stanton}, \citenamefont {Merer},\ and\ \citenamefont {Field}}]{Baraban2015}%
  \BibitemOpen
  \bibfield  {author} {\bibinfo {author} {\bibfnamefont {J.~H.}\ \bibnamefont {Baraban}}, \bibinfo {author} {\bibfnamefont {P.~B.}\ \bibnamefont {Changala}}, \bibinfo {author} {\bibfnamefont {G.~C.}\ \bibnamefont {Mellau}}, \bibinfo {author} {\bibfnamefont {J.~F.}\ \bibnamefont {Stanton}}, \bibinfo {author} {\bibfnamefont {A.~J.}\ \bibnamefont {Merer}}, \ and\ \bibinfo {author} {\bibfnamefont {R.~W.}\ \bibnamefont {Field}},\ }\bibfield  {title} {\enquote {\bibinfo {title} {Spectroscopic characterization of isomerization transition states},}\ }\href {\doibase 10.1126/science.aac9668} {\bibfield  {journal} {\bibinfo  {journal} {Science}\ }\textbf {\bibinfo {volume} {350}},\ \bibinfo {pages} {1338--1342} (\bibinfo {year} {2015})}\BibitemShut {NoStop}%
\bibitem [{\citenamefont {Eyring}\ and\ \citenamefont {Polanyi}(1931)}]{Eyring1931}%
  \BibitemOpen
  \bibfield  {author} {\bibinfo {author} {\bibfnamefont {H.}~\bibnamefont {Eyring}}\ and\ \bibinfo {author} {\bibfnamefont {M.}~\bibnamefont {Polanyi}},\ }\bibfield  {title} {\enquote {\bibinfo {title} {On simple gas reactions},}\ }\href@noop {} {\bibfield  {journal} {\bibinfo  {journal} {Z. Phys. Chem. B}\ }\textbf {\bibinfo {volume} {12}},\ \bibinfo {pages} {279} (\bibinfo {year} {1931})}\BibitemShut {NoStop}%
\bibitem [{\citenamefont {Nye}(2007)}]{Nye2007}%
  \BibitemOpen
  \bibfield  {author} {\bibinfo {author} {\bibfnamefont {M.~J.}\ \bibnamefont {Nye}},\ }\bibfield  {title} {\enquote {\bibinfo {title} {Working tools for theoretical chemistry: Polanyi, {Eyring}, and debates over the ‘‘semiempirical method’’},}\ }\href@noop {} {\bibfield  {journal} {\bibinfo  {journal} {J. Comput. Chem.}\ }\textbf {\bibinfo {volume} {28}},\ \bibinfo {pages} {98–108} (\bibinfo {year} {2007})}\BibitemShut {NoStop}%
\bibitem [{\citenamefont {Polanyi}\ and\ \citenamefont {Zewail}(1995)}]{Polanyi1995}%
  \BibitemOpen
  \bibfield  {author} {\bibinfo {author} {\bibfnamefont {J.~C.}\ \bibnamefont {Polanyi}}\ and\ \bibinfo {author} {\bibfnamefont {A.~H.}\ \bibnamefont {Zewail}},\ }\bibfield  {title} {\enquote {\bibinfo {title} {Direct observation of the transition state},}\ }\href@noop {} {\bibfield  {journal} {\bibinfo  {journal} {Acc. Chem. Res.}\ }\textbf {\bibinfo {volume} {28}},\ \bibinfo {pages} {119--132} (\bibinfo {year} {1995})}\BibitemShut {NoStop}%
\bibitem [{\citenamefont {Mellau}\ \emph {et~al.}(2016)\citenamefont {Mellau}, \citenamefont {Kyuberis}, \citenamefont {Polyansky}, \citenamefont {Zobov},\ and\ \citenamefont {Field}}]{mellau2016}%
  \BibitemOpen
  \bibfield  {author} {\bibinfo {author} {\bibfnamefont {G.~C.}\ \bibnamefont {Mellau}}, \bibinfo {author} {\bibfnamefont {A.~A.}\ \bibnamefont {Kyuberis}}, \bibinfo {author} {\bibfnamefont {O.~L.}\ \bibnamefont {Polyansky}}, \bibinfo {author} {\bibfnamefont {N.}~\bibnamefont {Zobov}}, \ and\ \bibinfo {author} {\bibfnamefont {R.~W.}\ \bibnamefont {Field}},\ }\bibfield  {title} {\enquote {\bibinfo {title} {Saddle point localization of molecular wavefunctions},}\ }\href {\doibase https://doi.org/10.1038/srep33068} {\bibfield  {journal} {\bibinfo  {journal} {Scientific Reports}\ }\textbf {\bibinfo {volume} {6}},\ \bibinfo {pages} {33068} (\bibinfo {year} {2016})}\BibitemShut {NoStop}%
\bibitem [{\citenamefont {Khalouf-Rivera}\ \emph {et~al.}(2019)\citenamefont {Khalouf-Rivera}, \citenamefont {Carvajal}, \citenamefont {Santos},\ and\ \citenamefont {P{\'e}rez-Bernal}}]{khalou2019}%
  \BibitemOpen
  \bibfield  {author} {\bibinfo {author} {\bibfnamefont {J.}~\bibnamefont {Khalouf-Rivera}}, \bibinfo {author} {\bibfnamefont {M.}~\bibnamefont {Carvajal}}, \bibinfo {author} {\bibfnamefont {L.~F.}\ \bibnamefont {Santos}}, \ and\ \bibinfo {author} {\bibfnamefont {F.}~\bibnamefont {P{\'e}rez-Bernal}},\ }\bibfield  {title} {\enquote {\bibinfo {title} {Calculation of transition state energies in the {HCN–HNC} isomerization with an algebraic model},}\ }\href {\doibase 10.1021/acs.jpca.9b07338} {\bibfield  {journal} {\bibinfo  {journal} {The Journal of Physical Chemistry A}\ }\textbf {\bibinfo {volume} {123}},\ \bibinfo {pages} {9544--9551} (\bibinfo {year} {2019})}\BibitemShut {NoStop}%
\bibitem [{\citenamefont {Maki}, \citenamefont {Plyler},\ and\ \citenamefont {Tidwell}(1962)}]{maki1962}%
  \BibitemOpen
  \bibfield  {author} {\bibinfo {author} {\bibfnamefont {A.~G.}\ \bibnamefont {Maki}}, \bibinfo {author} {\bibfnamefont {E.~K.}\ \bibnamefont {Plyler}}, \ and\ \bibinfo {author} {\bibfnamefont {E.~D.}\ \bibnamefont {Tidwell}},\ }\bibfield  {title} {\enquote {\bibinfo {title} {Vibration-rotation bands of carbonyl sulfide},}\ }\href@noop {} {\bibfield  {journal} {\bibinfo  {journal} {Journal of Research of the National Bureau of Standards. Section A, Physics and Chemistry}\ }\textbf {\bibinfo {volume} {66}},\ \bibinfo {pages} {163} (\bibinfo {year} {1962})}\BibitemShut {NoStop}%
\bibitem [{\citenamefont {Fayt}, \citenamefont {Vandenhaute},\ and\ \citenamefont {Lahaye}(1986)}]{fayt1986global}%
  \BibitemOpen
  \bibfield  {author} {\bibinfo {author} {\bibfnamefont {A.}~\bibnamefont {Fayt}}, \bibinfo {author} {\bibfnamefont {R.}~\bibnamefont {Vandenhaute}}, \ and\ \bibinfo {author} {\bibfnamefont {J.~G.}\ \bibnamefont {Lahaye}},\ }\bibfield  {title} {\enquote {\bibinfo {title} {Global rovibrational analysis of carbonyl sulfide},}\ }\href {\doibase https://doi.org/10.1016/0022-2852(86)90022-6} {\bibfield  {journal} {\bibinfo  {journal} {Journal of Molecular Spectroscopy}\ }\textbf {\bibinfo {volume} {119}},\ \bibinfo {pages} {233--266} (\bibinfo {year} {1986})}\BibitemShut {NoStop}%
\bibitem [{\citenamefont {Belafhal}, \citenamefont {Fayt},\ and\ \citenamefont {Guelachvili}(1995)}]{belafhal1995}%
  \BibitemOpen
  \bibfield  {author} {\bibinfo {author} {\bibfnamefont {A.}~\bibnamefont {Belafhal}}, \bibinfo {author} {\bibfnamefont {A.}~\bibnamefont {Fayt}}, \ and\ \bibinfo {author} {\bibfnamefont {G.}~\bibnamefont {Guelachvili}},\ }\bibfield  {title} {\enquote {\bibinfo {title} {Fourier {Transform} {Spectroscopy} of carbonyl sulfide from 1800 to 3120~cm$^{-1}$: The normal species},}\ }\href@noop {} {\bibfield  {journal} {\bibinfo  {journal} {J.\ Mol.\ Spectroscopy}\ }\textbf {\bibinfo {volume} {174}},\ \bibinfo {pages} {1--19} (\bibinfo {year} {1995})}\BibitemShut {NoStop}%
\bibitem [{\citenamefont {Hornberger}\ \emph {et~al.}(1996)\citenamefont {Hornberger}, \citenamefont {Boor}, \citenamefont {Stuber}, \citenamefont {Demtroder}, \citenamefont {Na\"im},\ and\ \citenamefont {Fayt}}]{hornberger1996}%
  \BibitemOpen
  \bibfield  {author} {\bibinfo {author} {\bibfnamefont {C.}~\bibnamefont {Hornberger}}, \bibinfo {author} {\bibfnamefont {B.}~\bibnamefont {Boor}}, \bibinfo {author} {\bibfnamefont {R.}~\bibnamefont {Stuber}}, \bibinfo {author} {\bibfnamefont {W.}~\bibnamefont {Demtroder}}, \bibinfo {author} {\bibfnamefont {S.}~\bibnamefont {Na\"im}}, \ and\ \bibinfo {author} {\bibfnamefont {A.}~\bibnamefont {Fayt}},\ }\bibfield  {title} {\enquote {\bibinfo {title} {Sensitive overtone spectroscopy of carbonyl sulfide between 6130 and 6650~cm$^{-1}$ and at 12000~cm$^{-1}$},}\ }\href@noop {} {\bibfield  {journal} {\bibinfo  {journal} {J.\ Mol.\ Spectroscopy}\ }\textbf {\bibinfo {volume} {179}},\ \bibinfo {pages} {237--245} (\bibinfo {year} {1996})}\BibitemShut {NoStop}%
\bibitem [{\citenamefont {Yang}\ and\ \citenamefont {Noda}(1997)}]{YANG1997}%
  \BibitemOpen
  \bibfield  {author} {\bibinfo {author} {\bibfnamefont {X.}~\bibnamefont {Yang}}\ and\ \bibinfo {author} {\bibfnamefont {C.}~\bibnamefont {Noda}},\ }\bibfield  {title} {\enquote {\bibinfo {title} {Vibrational overtone transitions of {OCS} in the near infrared},}\ }\href {\doibase https://doi.org/10.1006/jmsp.1997.7266} {\bibfield  {journal} {\bibinfo  {journal} {Journal of Molecular Spectroscopy}\ }\textbf {\bibinfo {volume} {183}},\ \bibinfo {pages} {151--156} (\bibinfo {year} {1997})}\BibitemShut {NoStop}%
\bibitem [{\citenamefont {Rbaihi}\ \emph {et~al.}(1998)\citenamefont {Rbaihi}, \citenamefont {Belafhal}, \citenamefont {Vander~Auwera}, \citenamefont {Na{\"i}m},\ and\ \citenamefont {Fayt}}]{rbaihi1998fourier}%
  \BibitemOpen
  \bibfield  {author} {\bibinfo {author} {\bibfnamefont {E.}~\bibnamefont {Rbaihi}}, \bibinfo {author} {\bibfnamefont {A.}~\bibnamefont {Belafhal}}, \bibinfo {author} {\bibfnamefont {J.}~\bibnamefont {Vander~Auwera}}, \bibinfo {author} {\bibfnamefont {S.}~\bibnamefont {Na{\"i}m}}, \ and\ \bibinfo {author} {\bibfnamefont {A.}~\bibnamefont {Fayt}},\ }\bibfield  {title} {\enquote {\bibinfo {title} {Fourier {Transform} {Spectroscopy} of carbonyl sulfide from 4800 to 8000~cm$^{-1}$ and new global analysis of $^{16}${O}$^{12}${C}$^{32}${S}},}\ }\href {\doibase https://doi.org/10.1006/jmsp.1998.7616} {\bibfield  {journal} {\bibinfo  {journal} {Journal of Molecular Spectroscopy}\ }\textbf {\bibinfo {volume} {191}},\ \bibinfo {pages} {32--44} (\bibinfo {year} {1998})}\BibitemShut {NoStop}%
\bibitem [{\citenamefont {Na\"im}\ \emph {et~al.}(1998)\citenamefont {Na\"im}, \citenamefont {Fayt}, \citenamefont {Bredohl}, \citenamefont {Blavier},\ and\ \citenamefont {Dubois}}]{naim1998}%
  \BibitemOpen
  \bibfield  {author} {\bibinfo {author} {\bibfnamefont {S.}~\bibnamefont {Na\"im}}, \bibinfo {author} {\bibfnamefont {A.}~\bibnamefont {Fayt}}, \bibinfo {author} {\bibfnamefont {H.}~\bibnamefont {Bredohl}}, \bibinfo {author} {\bibfnamefont {J.-F.}\ \bibnamefont {Blavier}}, \ and\ \bibinfo {author} {\bibfnamefont {I.}~\bibnamefont {Dubois}},\ }\bibfield  {title} {\enquote {\bibinfo {title} {Fourier transform spectroscopy of carbonyl sulfide from 3700 to 4800~cm$^{-1}$ and selection of a line-pointing program},}\ }\href@noop {} {\bibfield  {journal} {\bibinfo  {journal} {J.\ Mol.\ Spectroscopy}\ }\textbf {\bibinfo {volume} {192}},\ \bibinfo {pages} {91--101} (\bibinfo {year} {1998})}\BibitemShut {NoStop}%
\bibitem [{\citenamefont {Frech}\ \emph {et~al.}(1998)\citenamefont {Frech}, \citenamefont {Murtz}, \citenamefont {Palm}, \citenamefont {Lotze}, \citenamefont {Urban},\ and\ \citenamefont {Maki}}]{frech1998}%
  \BibitemOpen
  \bibfield  {author} {\bibinfo {author} {\bibfnamefont {B.}~\bibnamefont {Frech}}, \bibinfo {author} {\bibfnamefont {M.}~\bibnamefont {Murtz}}, \bibinfo {author} {\bibfnamefont {P.}~\bibnamefont {Palm}}, \bibinfo {author} {\bibfnamefont {R.}~\bibnamefont {Lotze}}, \bibinfo {author} {\bibfnamefont {W.}~\bibnamefont {Urban}}, \ and\ \bibinfo {author} {\bibfnamefont {A.}~\bibnamefont {Maki}},\ }\bibfield  {title} {\enquote {\bibinfo {title} {{Sub-Doppler} {Heterodyne} frequency measurements on {OCS} near 2900~cm$^{-1}$ using a co overtone sideband spectrometer},}\ }\href@noop {} {\bibfield  {journal} {\bibinfo  {journal} {J.\ Mol.\ Spectroscopy}\ }\textbf {\bibinfo {volume} {190}},\ \bibinfo {pages} {91--100} (\bibinfo {year} {1998})}\BibitemShut {NoStop}%
\bibitem [{\citenamefont {Tranchart}\ \emph {et~al.}(1999)\citenamefont {Tranchart}, \citenamefont {{Hadj Bachir}}, \citenamefont {Huet}, \citenamefont {Olafsson}, \citenamefont {Destombes}, \citenamefont {Naı¨m},\ and\ \citenamefont {Fayt}}]{TRANCHART1999}%
  \BibitemOpen
  \bibfield  {author} {\bibinfo {author} {\bibfnamefont {S.}~\bibnamefont {Tranchart}}, \bibinfo {author} {\bibfnamefont {I.}~\bibnamefont {{Hadj Bachir}}}, \bibinfo {author} {\bibfnamefont {T.}~\bibnamefont {Huet}}, \bibinfo {author} {\bibfnamefont {A.}~\bibnamefont {Olafsson}}, \bibinfo {author} {\bibfnamefont {J.-L.}\ \bibnamefont {Destombes}}, \bibinfo {author} {\bibfnamefont {S.}~\bibnamefont {Naı¨m}}, \ and\ \bibinfo {author} {\bibfnamefont {A.}~\bibnamefont {Fayt}},\ }\bibfield  {title} {\enquote {\bibinfo {title} {High-resolution laser photoacoustic spectroscopy of ocs in the 12000–13000 cm$^{-1}$ region},}\ }\href {\doibase https://doi.org/10.1006/jmsp.1999.7853} {\bibfield  {journal} {\bibinfo  {journal} {Journal of Molecular Spectroscopy}\ }\textbf {\bibinfo {volume} {196}},\ \bibinfo {pages} {265--273} (\bibinfo {year} {1999})}\BibitemShut {NoStop}%
\bibitem [{\citenamefont {Golebiowski}\ \emph {et~al.}(2014)\citenamefont {Golebiowski}, \citenamefont {{de Ghellinck d’Elseghem Vaernewijck}}, \citenamefont {Herman}, \citenamefont {{Vander Auwera}},\ and\ \citenamefont {Fayt}}]{GOLEBIOWSKI2014}%
  \BibitemOpen
  \bibfield  {author} {\bibinfo {author} {\bibfnamefont {D.}~\bibnamefont {Golebiowski}}, \bibinfo {author} {\bibfnamefont {X.}~\bibnamefont {{de Ghellinck d’Elseghem Vaernewijck}}}, \bibinfo {author} {\bibfnamefont {M.}~\bibnamefont {Herman}}, \bibinfo {author} {\bibfnamefont {J.}~\bibnamefont {{Vander Auwera}}}, \ and\ \bibinfo {author} {\bibfnamefont {A.}~\bibnamefont {Fayt}},\ }\bibfield  {title} {\enquote {\bibinfo {title} {High sensitivity ({femto-FT-CEAS}) spectra of carbonyl sulphide between 6200 and 8200cm$^{-1}$, and new energy pattern in the global rovibrational analysis of {$^{16}$O$^{12}$C$^{32}$S}},}\ }\href {\doibase https://doi.org/10.1016/j.jqsrt.2014.07.005} {\bibfield  {journal} {\bibinfo  {journal} {Journal of Quantitative Spectroscopy and Radiative Transfer}\ }\textbf {\bibinfo {volume} {149}},\ \bibinfo {pages} {184--203} (\bibinfo {year} {2014})}\BibitemShut {NoStop}%
\bibitem [{\citenamefont {Br{\"u}hl}\ \emph {et~al.}(2012)\citenamefont {Br{\"u}hl}, \citenamefont {Lelieveld}, \citenamefont {Crutzen},\ and\ \citenamefont {Tost}}]{bruhl2012}%
  \BibitemOpen
  \bibfield  {author} {\bibinfo {author} {\bibfnamefont {C.}~\bibnamefont {Br{\"u}hl}}, \bibinfo {author} {\bibfnamefont {J.}~\bibnamefont {Lelieveld}}, \bibinfo {author} {\bibfnamefont {P.~J.}\ \bibnamefont {Crutzen}}, \ and\ \bibinfo {author} {\bibfnamefont {H.}~\bibnamefont {Tost}},\ }\bibfield  {title} {\enquote {\bibinfo {title} {The role of carbonyl sulphide as a source of stratospheric sulphate aerosol and its impact on climate},}\ }\href {\doibase https://doi.org/10.5194/acp-12-1239-2012} {\bibfield  {journal} {\bibinfo  {journal} {Atmospheric Chemistry and Physics}\ }\textbf {\bibinfo {volume} {12}},\ \bibinfo {pages} {1239--1253} (\bibinfo {year} {2012})}\BibitemShut {NoStop}%
\bibitem [{\citenamefont {Ma}\ \emph {et~al.}(2021)\citenamefont {Ma}, \citenamefont {Kooijmans}, \citenamefont {Cho}, \citenamefont {Montzka}, \citenamefont {Glatthor}, \citenamefont {Worden}, \citenamefont {Kuai}, \citenamefont {Atlas},\ and\ \citenamefont {Krol}}]{ma2021}%
  \BibitemOpen
  \bibfield  {author} {\bibinfo {author} {\bibfnamefont {J.}~\bibnamefont {Ma}}, \bibinfo {author} {\bibfnamefont {L.~M.~J.}\ \bibnamefont {Kooijmans}}, \bibinfo {author} {\bibfnamefont {A.}~\bibnamefont {Cho}}, \bibinfo {author} {\bibfnamefont {S.~A.}\ \bibnamefont {Montzka}}, \bibinfo {author} {\bibfnamefont {N.}~\bibnamefont {Glatthor}}, \bibinfo {author} {\bibfnamefont {J.~R.}\ \bibnamefont {Worden}}, \bibinfo {author} {\bibfnamefont {L.}~\bibnamefont {Kuai}}, \bibinfo {author} {\bibfnamefont {E.~L.}\ \bibnamefont {Atlas}}, \ and\ \bibinfo {author} {\bibfnamefont {M.~C.}\ \bibnamefont {Krol}},\ }\bibfield  {title} {\enquote {\bibinfo {title} {Inverse modelling of carbonyl sulfide: implementation, evaluation and implications for the global budget},}\ }\href {\doibase https://doi.org/10.5194/acp-21-3507-2021} {\bibfield  {journal} {\bibinfo  {journal} {Atmospheric Chemistry and Physics}\ }\textbf {\bibinfo {volume} {21}},\ \bibinfo {pages} {3507--3529} (\bibinfo {year} {2021})}\BibitemShut {NoStop}%
\bibitem [{\citenamefont {Campbell}\ \emph {et~al.}(2008)\citenamefont {Campbell}, \citenamefont {Carmichael}, \citenamefont {Chai}, \citenamefont {Mena-Carrasco}, \citenamefont {Tang}, \citenamefont {Blake}, \citenamefont {Blake}, \citenamefont {Vay}, \citenamefont {Collatz}, \citenamefont {Baker} \emph {et~al.}}]{campbell2008}%
  \BibitemOpen
  \bibfield  {author} {\bibinfo {author} {\bibfnamefont {J.~E.}\ \bibnamefont {Campbell}}, \bibinfo {author} {\bibfnamefont {G.~R.}\ \bibnamefont {Carmichael}}, \bibinfo {author} {\bibfnamefont {T.}~\bibnamefont {Chai}}, \bibinfo {author} {\bibfnamefont {M.}~\bibnamefont {Mena-Carrasco}}, \bibinfo {author} {\bibfnamefont {Y.}~\bibnamefont {Tang}}, \bibinfo {author} {\bibfnamefont {D.}~\bibnamefont {Blake}}, \bibinfo {author} {\bibfnamefont {N.}~\bibnamefont {Blake}}, \bibinfo {author} {\bibfnamefont {S.~A.}\ \bibnamefont {Vay}}, \bibinfo {author} {\bibfnamefont {G.~J.}\ \bibnamefont {Collatz}}, \bibinfo {author} {\bibfnamefont {I.}~\bibnamefont {Baker}},  \emph {et~al.},\ }\bibfield  {title} {\enquote {\bibinfo {title} {Photosynthetic control of atmospheric carbonyl sulfide during the growing season},}\ }\href {\doibase 10.1126/science.1164015} {\bibfield  {journal} {\bibinfo  {journal} {Science}\ }\textbf {\bibinfo {volume} {322}},\ \bibinfo {pages} {1085--1088} (\bibinfo {year} {2008})}\BibitemShut
  {NoStop}%
\bibitem [{\citenamefont {De~Gouw}\ \emph {et~al.}(2009)\citenamefont {De~Gouw}, \citenamefont {Warneke}, \citenamefont {Montzka}, \citenamefont {Holloway}, \citenamefont {Parrish}, \citenamefont {Fehsenfeld}, \citenamefont {Atlas}, \citenamefont {Weber},\ and\ \citenamefont {Flocke}}]{DeGouw2009}%
  \BibitemOpen
  \bibfield  {author} {\bibinfo {author} {\bibfnamefont {J.~A.}\ \bibnamefont {De~Gouw}}, \bibinfo {author} {\bibfnamefont {C.}~\bibnamefont {Warneke}}, \bibinfo {author} {\bibfnamefont {S.~A.}\ \bibnamefont {Montzka}}, \bibinfo {author} {\bibfnamefont {J.~S.}\ \bibnamefont {Holloway}}, \bibinfo {author} {\bibfnamefont {D.~D.}\ \bibnamefont {Parrish}}, \bibinfo {author} {\bibfnamefont {F.~C.}\ \bibnamefont {Fehsenfeld}}, \bibinfo {author} {\bibfnamefont {E.~L.}\ \bibnamefont {Atlas}}, \bibinfo {author} {\bibfnamefont {R.~J.}\ \bibnamefont {Weber}}, \ and\ \bibinfo {author} {\bibfnamefont {F.~M.}\ \bibnamefont {Flocke}},\ }\bibfield  {title} {\enquote {\bibinfo {title} {Carbonyl sulfide as an inverse tracer for biogenic organic carbon in gas and aerosol phases},}\ }\href@noop {} {\bibfield  {journal} {\bibinfo  {journal} {Geophysical research letters}\ }\textbf {\bibinfo {volume} {36}},\ \bibinfo {pages} {L05804} (\bibinfo {year} {2009})}\BibitemShut {NoStop}%
\bibitem [{\citenamefont {Lellouch}\ \emph {et~al.}(1995)\citenamefont {Lellouch}, \citenamefont {Paubert}, \citenamefont {Moreno}, \citenamefont {Festou}, \citenamefont {B{\'e}zard}, \citenamefont {Bockel{\'e}e-Morvan}, \citenamefont {Colom}, \citenamefont {Crovisier}, \citenamefont {Encrenaz}, \citenamefont {Gautier} \emph {et~al.}}]{lellouch1995}%
  \BibitemOpen
  \bibfield  {author} {\bibinfo {author} {\bibfnamefont {E.}~\bibnamefont {Lellouch}}, \bibinfo {author} {\bibfnamefont {G.}~\bibnamefont {Paubert}}, \bibinfo {author} {\bibfnamefont {R.}~\bibnamefont {Moreno}}, \bibinfo {author} {\bibfnamefont {M.~C.}\ \bibnamefont {Festou}}, \bibinfo {author} {\bibfnamefont {B.}~\bibnamefont {B{\'e}zard}}, \bibinfo {author} {\bibfnamefont {D.}~\bibnamefont {Bockel{\'e}e-Morvan}}, \bibinfo {author} {\bibfnamefont {P.}~\bibnamefont {Colom}}, \bibinfo {author} {\bibfnamefont {J.}~\bibnamefont {Crovisier}}, \bibinfo {author} {\bibfnamefont {T.}~\bibnamefont {Encrenaz}}, \bibinfo {author} {\bibfnamefont {D.}~\bibnamefont {Gautier}},  \emph {et~al.},\ }\bibfield  {title} {\enquote {\bibinfo {title} {Chemical and thermal response of jupiter's atmosphere following the impact of comet {Shoemaker--Levy} 9},}\ }\href {\doibase https://doi.org/10.1038/373592a0} {\bibfield  {journal} {\bibinfo  {journal} {Nature}\ }\textbf {\bibinfo {volume} {373}},\ \bibinfo {pages} {592--595}
  (\bibinfo {year} {1995})}\BibitemShut {NoStop}%
\bibitem [{\citenamefont {B{\'e}zard}\ \emph {et~al.}(1990)\citenamefont {B{\'e}zard}, \citenamefont {De~Bergh}, \citenamefont {Crisp},\ and\ \citenamefont {Maillard}}]{bezard1990}%
  \BibitemOpen
  \bibfield  {author} {\bibinfo {author} {\bibfnamefont {B.}~\bibnamefont {B{\'e}zard}}, \bibinfo {author} {\bibfnamefont {C.}~\bibnamefont {De~Bergh}}, \bibinfo {author} {\bibfnamefont {D.}~\bibnamefont {Crisp}}, \ and\ \bibinfo {author} {\bibfnamefont {J.-p.}\ \bibnamefont {Maillard}},\ }\bibfield  {title} {\enquote {\bibinfo {title} {The deep atmosphere of venus revealed by high-resolution nightside spectra},}\ }\href {\doibase https://doi.org/10.1038/345508a0} {\bibfield  {journal} {\bibinfo  {journal} {Nature}\ }\textbf {\bibinfo {volume} {345}},\ \bibinfo {pages} {508--511} (\bibinfo {year} {1990})}\BibitemShut {NoStop}%
\bibitem [{\citenamefont {Jefferts}\ \emph {et~al.}(1971)\citenamefont {Jefferts}, \citenamefont {Penzias}, \citenamefont {Wilson},\ and\ \citenamefont {Solomon}}]{jefferts1971}%
  \BibitemOpen
  \bibfield  {author} {\bibinfo {author} {\bibfnamefont {K.~B.}\ \bibnamefont {Jefferts}}, \bibinfo {author} {\bibfnamefont {A.~A.}\ \bibnamefont {Penzias}}, \bibinfo {author} {\bibfnamefont {R.~W.}\ \bibnamefont {Wilson}}, \ and\ \bibinfo {author} {\bibfnamefont {P.~M.}\ \bibnamefont {Solomon}},\ }\bibfield  {title} {\enquote {\bibinfo {title} {Detection of interstellar carbonyl sulfide},}\ }\href {\doibase https://doi.org/10.1086/180795} {\bibfield  {journal} {\bibinfo  {journal} {Astrophysical Journal}\ }\textbf {\bibinfo {volume} {168}},\ \bibinfo {pages} {L111--L113} (\bibinfo {year} {1971})}\BibitemShut {NoStop}%
\bibitem [{\citenamefont {Mauersberger}, \citenamefont {Henkel},\ and\ \citenamefont {Chin}(1995)}]{mauersberger1995}%
  \BibitemOpen
  \bibfield  {author} {\bibinfo {author} {\bibfnamefont {R.}~\bibnamefont {Mauersberger}}, \bibinfo {author} {\bibfnamefont {C.}~\bibnamefont {Henkel}}, \ and\ \bibinfo {author} {\bibfnamefont {Y.-N.}\ \bibnamefont {Chin}},\ }\bibfield  {title} {\enquote {\bibinfo {title} {Dense gas in nearby galaxies. viii. the detection of {OCS}},}\ }\href@noop {} {\bibfield  {journal} {\bibinfo  {journal} {Astronomy and Astrophysics}\ }\textbf {\bibinfo {volume} {294}},\ \bibinfo {pages} {23--32} (\bibinfo {year} {1995})}\BibitemShut {NoStop}%
\bibitem [{\citenamefont {Zúñiga}\ \emph {et~al.}(2000)\citenamefont {Zúñiga}, \citenamefont {Bastida}, \citenamefont {Alacid},\ and\ \citenamefont {Requena}}]{zuniga2000}%
  \BibitemOpen
  \bibfield  {author} {\bibinfo {author} {\bibfnamefont {J.}~\bibnamefont {Zúñiga}}, \bibinfo {author} {\bibfnamefont {A.}~\bibnamefont {Bastida}}, \bibinfo {author} {\bibfnamefont {M.}~\bibnamefont {Alacid}}, \ and\ \bibinfo {author} {\bibfnamefont {A.}~\bibnamefont {Requena}},\ }\bibfield  {title} {\enquote {\bibinfo {title} {Excited vibrational states and potential energy function for {OCS} determined using generalized internal coordinates},}\ }\href@noop {} {\bibfield  {journal} {\bibinfo  {journal} {The Journal of Chemical Physics}\ }\textbf {\bibinfo {volume} {113}},\ \bibinfo {pages} {5695--5704} (\bibinfo {year} {2000})}\BibitemShut {NoStop}%
\bibitem [{\citenamefont {Xie}\ \emph {et~al.}(2001)\citenamefont {Xie}, \citenamefont {Lu}, \citenamefont {Xu},\ and\ \citenamefont {Yan}}]{xie2001}%
  \BibitemOpen
  \bibfield  {author} {\bibinfo {author} {\bibfnamefont {D.}~\bibnamefont {Xie}}, \bibinfo {author} {\bibfnamefont {Y.}~\bibnamefont {Lu}}, \bibinfo {author} {\bibfnamefont {D.}~\bibnamefont {Xu}}, \ and\ \bibinfo {author} {\bibfnamefont {G.}~\bibnamefont {Yan}},\ }\bibfield  {title} {\enquote {\bibinfo {title} {Theoretical studies on the potential energy surface and rovibrational states for the electronic ground state of carbonyl sulfide},}\ }\href {\doibase https://doi.org/10.1016/S0301-0104(01)00406-2} {\bibfield  {journal} {\bibinfo  {journal} {Chemical Physics}\ }\textbf {\bibinfo {volume} {270}},\ \bibinfo {pages} {415--428} (\bibinfo {year} {2001})}\BibitemShut {NoStop}%
\bibitem [{\citenamefont {Sedivcov{\'a}-Uhl{\'\i}kov{\'a}}, \citenamefont {Abdullah},\ and\ \citenamefont {Manini}(2009)}]{sedivcova2009}%
  \BibitemOpen
  \bibfield  {author} {\bibinfo {author} {\bibfnamefont {T.}~\bibnamefont {Sedivcov{\'a}-Uhl{\'\i}kov{\'a}}}, \bibinfo {author} {\bibfnamefont {H.~Y.}\ \bibnamefont {Abdullah}}, \ and\ \bibinfo {author} {\bibfnamefont {N.}~\bibnamefont {Manini}},\ }\bibfield  {title} {\enquote {\bibinfo {title} {Algebraic-matrix calculation of vibrational levels of triatomic molecules},}\ }\href {\doibase 10.1021/jp8105474} {\bibfield  {journal} {\bibinfo  {journal} {The Journal of Physical Chemistry A}\ }\textbf {\bibinfo {volume} {113}},\ \bibinfo {pages} {6142--6148} (\bibinfo {year} {2009})}\BibitemShut {NoStop}%
\bibitem [{\citenamefont {Su\'arez}, \citenamefont {Guzm\'an-Ju\'arez},\ and\ \citenamefont {Lemus}(2025)}]{Suarez2025}%
  \BibitemOpen
  \bibfield  {author} {\bibinfo {author} {\bibfnamefont {E.}~\bibnamefont {Su\'arez}}, \bibinfo {author} {\bibfnamefont {O.}~\bibnamefont {Guzm\'an-Ju\'arez}}, \ and\ \bibinfo {author} {\bibfnamefont {R.}~\bibnamefont {Lemus}},\ }\bibfield  {title} {\enquote {\bibinfo {title} {Description of vibrational excitations of the {OCS} molecule using a local algebraic approach},}\ }\href@noop {} {\bibfield  {journal} {\bibinfo  {journal} {J.\ Quant.\ Spectr.\ Rad.\ Transf.}\ }\textbf {\bibinfo {volume} {340}},\ \bibinfo {pages} {109432} (\bibinfo {year} {2025})}\BibitemShut {NoStop}%
\bibitem [{\citenamefont {Xu}\ and\ \citenamefont {Tennyson}(2024)}]{xu2024empirical}%
  \BibitemOpen
  \bibfield  {author} {\bibinfo {author} {\bibfnamefont {E.}~\bibnamefont {Xu}}\ and\ \bibinfo {author} {\bibfnamefont {J.}~\bibnamefont {Tennyson}},\ }\bibfield  {title} {\enquote {\bibinfo {title} {Empirical rovibrational energy levels for carbonyl sulphide},}\ }\href {\doibase 10.1080/00268976.2023.2279694} {\bibfield  {journal} {\bibinfo  {journal} {Molecular Physics}\ }\textbf {\bibinfo {volume} {122}},\ \bibinfo {pages} {e2279694} (\bibinfo {year} {2024})}\BibitemShut {NoStop}%
\bibitem [{\citenamefont {Dobrolyubov}\ \emph {et~al.}(2024)\citenamefont {Dobrolyubov}, \citenamefont {Polyakov}, \citenamefont {Millionshchikov},\ and\ \citenamefont {Krasnoshchekov}}]{Dobrolyubov2024}%
  \BibitemOpen
  \bibfield  {author} {\bibinfo {author} {\bibfnamefont {E.~O.}\ \bibnamefont {Dobrolyubov}}, \bibinfo {author} {\bibfnamefont {I.~V.}\ \bibnamefont {Polyakov}}, \bibinfo {author} {\bibfnamefont {D.~V.}\ \bibnamefont {Millionshchikov}}, \ and\ \bibinfo {author} {\bibfnamefont {S.~V.}\ \bibnamefont {Krasnoshchekov}},\ }\bibfield  {title} {\enquote {\bibinfo {title} {Vibrational resonance phenomena of the {OCS} isotopologues studied by resummation of high-order {Rayleigh--Schr{\"o}dinger} perturbation theory},}\ }\href {\doibase 0.1016/j.jqsrt.2024.108909} {\bibfield  {journal} {\bibinfo  {journal} {Journal of Quantitative Spectroscopy and Radiative Transfer}\ }\textbf {\bibinfo {volume} {316}},\ \bibinfo {pages} {108909} (\bibinfo {year} {2024})}\BibitemShut {NoStop}%
\bibitem [{\citenamefont {Owens}(2024)}]{owens2024abinitio}%
  \BibitemOpen
  \bibfield  {author} {\bibinfo {author} {\bibfnamefont {A.}~\bibnamefont {Owens}},\ }\bibfield  {title} {\enquote {\bibinfo {title} {A highly accurate potential energy surface for carbonyl sulphide ({OCS}): how important are the ab initio calculations?}}\ }\href {\doibase 10.1039/D4CP01205D} {\bibfield  {journal} {\bibinfo  {journal} {Phys. Chem. Chem. Phys.}\ }\textbf {\bibinfo {volume} {26}},\ \bibinfo {pages} {17684--17694} (\bibinfo {year} {2024})}\BibitemShut {NoStop}%
\bibitem [{\citenamefont {Owens}, \citenamefont {Yurchenko},\ and\ \citenamefont {Tennyson}(2024)}]{owens2024exomol}%
  \BibitemOpen
  \bibfield  {author} {\bibinfo {author} {\bibfnamefont {A.}~\bibnamefont {Owens}}, \bibinfo {author} {\bibfnamefont {S.~N.}\ \bibnamefont {Yurchenko}}, \ and\ \bibinfo {author} {\bibfnamefont {J.}~\bibnamefont {Tennyson}},\ }\bibfield  {title} {\enquote {\bibinfo {title} {Exomol line lists--lviii. high-temperature molecular line list of carbonyl sulphide ({OCS})},}\ }\href {\doibase 10.1093/mnras/stae1110} {\bibfield  {journal} {\bibinfo  {journal} {Monthly Notices of the Royal Astronomical Society}\ }\textbf {\bibinfo {volume} {530}},\ \bibinfo {pages} {4004--4015} (\bibinfo {year} {2024})}\BibitemShut {NoStop}%
\bibitem [{\citenamefont {Huang}\ \emph {et~al.}(2025)\citenamefont {Huang}, \citenamefont {Gordon}, \citenamefont {Bertin}, \citenamefont {Schwenke},\ and\ \citenamefont {Lee}}]{Huang2025}%
  \BibitemOpen
  \bibfield  {author} {\bibinfo {author} {\bibfnamefont {X.}~\bibnamefont {Huang}}, \bibinfo {author} {\bibfnamefont {I.~E.}\ \bibnamefont {Gordon}}, \bibinfo {author} {\bibfnamefont {T.}~\bibnamefont {Bertin}}, \bibinfo {author} {\bibfnamefont {D.~W.}\ \bibnamefont {Schwenke}}, \ and\ \bibinfo {author} {\bibfnamefont {T.~J.}\ \bibnamefont {Lee}},\ }\bibfield  {title} {\enquote {\bibinfo {title} {Accurate potential energy surface, dipole moment surface, and ir line lists for {OCS} isotopologues up to 2000~{K}},}\ }\href {\doibase https://doi.org/10.1016/j.jqsrt.2025.109425} {\bibfield  {journal} {\bibinfo  {journal} {Journal of Quantitative Spectroscopy and Radiative Transfer}\ ,\ \bibinfo {pages} {109425}} (\bibinfo {year} {2025})}\BibitemShut {NoStop}%
\bibitem [{\citenamefont {Lara-Cruz}\ and\ \citenamefont {Moyano}(2018)}]{lara2018}%
  \BibitemOpen
  \bibfield  {author} {\bibinfo {author} {\bibfnamefont {G.~A.}\ \bibnamefont {Lara-Cruz}}\ and\ \bibinfo {author} {\bibfnamefont {G.~E.}\ \bibnamefont {Moyano}},\ }\bibfield  {title} {\enquote {\bibinfo {title} {{OCS} isomerization and dissociation kinetics from statistical models},}\ }\href {\doibase 10.1007/s00214-018-2253-9} {\bibfield  {journal} {\bibinfo  {journal} {Theoretical Chemistry Accounts}\ }\textbf {\bibinfo {volume} {137}},\ \bibinfo {pages} {79} (\bibinfo {year} {2018})}\BibitemShut {NoStop}%
\bibitem [{\citenamefont {Iachello}(2015)}]{bookalg}%
  \BibitemOpen
  \bibfield  {author} {\bibinfo {author} {\bibfnamefont {F.}~\bibnamefont {Iachello}},\ }\href@noop {} {\emph {\bibinfo {title} {Lie Algebras and Applications (Lecture Notes in Physics)}}},\ Vol.\ \bibinfo {volume} {891}\ (\bibinfo  {publisher} {Springer, Berlin},\ \bibinfo {year} {2015})\BibitemShut {NoStop}%
\bibitem [{\citenamefont {Iachello}(1981)}]{iachello1981}%
  \BibitemOpen
  \bibfield  {author} {\bibinfo {author} {\bibfnamefont {F.}~\bibnamefont {Iachello}},\ }\bibfield  {title} {\enquote {\bibinfo {title} {Algebraic methods for molecular rotation-vibration spectra},}\ }\href {\doibase http://dx.doi.org/10.1016/0009-2614(81)85262-1} {\bibfield  {journal} {\bibinfo  {journal} {Chemical Physics Letters}\ }\textbf {\bibinfo {volume} {78}},\ \bibinfo {pages} {581--585} (\bibinfo {year} {1981})}\BibitemShut {NoStop}%
\bibitem [{\citenamefont {Iachello}\ and\ \citenamefont {Levine}(1995)}]{bookmol}%
  \BibitemOpen
  \bibfield  {author} {\bibinfo {author} {\bibfnamefont {F.}~\bibnamefont {Iachello}}\ and\ \bibinfo {author} {\bibfnamefont {R.~D.}\ \bibnamefont {Levine}},\ }\href@noop {} {\emph {\bibinfo {title} {Algebraic Theory of Molecules}}}\ (\bibinfo  {publisher} {Oxford University Press, Oxford},\ \bibinfo {year} {1995})\BibitemShut {NoStop}%
\bibitem [{\citenamefont {Frank}\ and\ \citenamefont {Isacker}(1994)}]{frank}%
  \BibitemOpen
  \bibfield  {author} {\bibinfo {author} {\bibfnamefont {A.}~\bibnamefont {Frank}}\ and\ \bibinfo {author} {\bibfnamefont {P.~V.}\ \bibnamefont {Isacker}},\ }\href@noop {} {\emph {\bibinfo {title} {Algebraic Methods in Molecular and Nuclear Structure Physics}}}\ (\bibinfo  {publisher} {John Wiley and Sons, New York},\ \bibinfo {year} {1994})\BibitemShut {NoStop}%
\bibitem [{\citenamefont {Oss}(1996)}]{Oss1996}%
  \BibitemOpen
  \bibfield  {author} {\bibinfo {author} {\bibfnamefont {S.}~\bibnamefont {Oss}},\ }\enquote {\bibinfo {title} {Algebraic models in molecular spectroscopy},}\ in\ \href {\doibase https://doi.org/10.1002/9780470141526.ch8} {\emph {\bibinfo {booktitle} {Adv. Chem. Phys.}}}\ (\bibinfo  {publisher} {John Wiley \& Sons, Ltd},\ \bibinfo {year} {1996})\ pp.\ \bibinfo {pages} {455--649}\BibitemShut {NoStop}%
\bibitem [{\citenamefont {Iachello}\ and\ \citenamefont {Oss}(1996)}]{iachello1996}%
  \BibitemOpen
  \bibfield  {author} {\bibinfo {author} {\bibfnamefont {F.}~\bibnamefont {Iachello}}\ and\ \bibinfo {author} {\bibfnamefont {S.}~\bibnamefont {Oss}},\ }\bibfield  {title} {\enquote {\bibinfo {title} {Algebraic approach to molecular spectra: Two-dimensional problems},}\ }\href {\doibase https://doi.org/10.1063/1.471412} {\bibfield  {journal} {\bibinfo  {journal} {The Journal of chemical physics}\ }\textbf {\bibinfo {volume} {104}},\ \bibinfo {pages} {6956--6963} (\bibinfo {year} {1996})}\BibitemShut {NoStop}%
\bibitem [{\citenamefont {P{\'e}rez-Bernal}\ and\ \citenamefont {Iachello}(2008)}]{perez2008}%
  \BibitemOpen
  \bibfield  {author} {\bibinfo {author} {\bibfnamefont {F.}~\bibnamefont {P{\'e}rez-Bernal}}\ and\ \bibinfo {author} {\bibfnamefont {F.}~\bibnamefont {Iachello}},\ }\bibfield  {title} {\enquote {\bibinfo {title} {Algebraic approach to two-dimensional systems: Shape phase transitions, monodromy, and thermodynamic quantities},}\ }\href {\doibase https://doi.org/10.1103/PhysRevA.77.032115} {\bibfield  {journal} {\bibinfo  {journal} {Physical Review A—Atomic, Molecular, and Optical Physics}\ }\textbf {\bibinfo {volume} {77}},\ \bibinfo {pages} {032115} (\bibinfo {year} {2008})}\BibitemShut {NoStop}%
\bibitem [{\citenamefont {Iachello}, \citenamefont {Pérez-Bernal},\ and\ \citenamefont {Vaccaro}(2003)}]{iachello2003}%
  \BibitemOpen
  \bibfield  {author} {\bibinfo {author} {\bibfnamefont {F.}~\bibnamefont {Iachello}}, \bibinfo {author} {\bibfnamefont {F.}~\bibnamefont {Pérez-Bernal}}, \ and\ \bibinfo {author} {\bibfnamefont {P.}~\bibnamefont {Vaccaro}},\ }\bibfield  {title} {\enquote {\bibinfo {title} {A novel algebraic scheme for describing nonrigid molecules},}\ }\href@noop {} {\bibfield  {journal} {\bibinfo  {journal} {Chemical Physics Letters}\ }\textbf {\bibinfo {volume} {375}},\ \bibinfo {pages} {309--20} (\bibinfo {year} {2003})}\BibitemShut {NoStop}%
\bibitem [{\citenamefont {Pérez-Bernal}\ \emph {et~al.}(2005)\citenamefont {Pérez-Bernal}, \citenamefont {Santos}, \citenamefont {Vaccaro},\ and\ \citenamefont {Iachello}}]{PBernal2005}%
  \BibitemOpen
  \bibfield  {author} {\bibinfo {author} {\bibfnamefont {F.}~\bibnamefont {Pérez-Bernal}}, \bibinfo {author} {\bibfnamefont {L.~F.}\ \bibnamefont {Santos}}, \bibinfo {author} {\bibfnamefont {P.~H.}\ \bibnamefont {Vaccaro}}, \ and\ \bibinfo {author} {\bibfnamefont {F.}~\bibnamefont {Iachello}},\ }\bibfield  {title} {\enquote {\bibinfo {title} {Spectroscopic signatures of nonrigidity: Algebraic analyses of infrared and {Raman} transitions in nonrigid species},}\ }\href {\doibase https://doi.org/10.1016/j.cplett.2005.07.119} {\bibfield  {journal} {\bibinfo  {journal} {Chem.\ Phys.\ Lett.}\ }\textbf {\bibinfo {volume} {414}},\ \bibinfo {pages} {398 -- 404} (\bibinfo {year} {2005})}\BibitemShut {NoStop}%
\bibitem [{\citenamefont {Dixon}(1964)}]{Dixon1964}%
  \BibitemOpen
  \bibfield  {author} {\bibinfo {author} {\bibfnamefont {R.~N.}\ \bibnamefont {Dixon}},\ }\bibfield  {title} {\enquote {\bibinfo {title} {{H}igher {V}ibrational {L}evels of a {B}ent {T}riatomic {M}olecule},}\ }\href {\doibase 10.1039/TF9646001363} {\bibfield  {journal} {\bibinfo  {journal} {Trans. Faraday Soc.}\ }\textbf {\bibinfo {volume} {60}},\ \bibinfo {pages} {1363--1368} (\bibinfo {year} {1964})}\BibitemShut {NoStop}%
\bibitem [{\citenamefont {Yamada}\ and\ \citenamefont {Winnewisser}(1976)}]{Yamada1976}%
  \BibitemOpen
  \bibfield  {author} {\bibinfo {author} {\bibfnamefont {K.}~\bibnamefont {Yamada}}\ and\ \bibinfo {author} {\bibfnamefont {M.}~\bibnamefont {Winnewisser}},\ }\bibfield  {title} {\enquote {\bibinfo {title} {{A Parameter to Quantify Molecular Quasilinearity}},}\ }\href {\doibase https://doi.org/10.1515/zna-1976-0206} {\bibfield  {journal} {\bibinfo  {journal} {Z. Naturforsch. A}\ }\textbf {\bibinfo {volume} {31}},\ \bibinfo {pages} {139 -- 144} (\bibinfo {year} {1976})}\BibitemShut {NoStop}%
\bibitem [{\citenamefont {Child}(1998)}]{Child1998}%
  \BibitemOpen
  \bibfield  {author} {\bibinfo {author} {\bibfnamefont {M.~S.}\ \bibnamefont {Child}},\ }\bibfield  {title} {\enquote {\bibinfo {title} {Quantum states in a champagne bottle},}\ }\href {\doibase 10.1088/0305-4470/31/2/022} {\bibfield  {journal} {\bibinfo  {journal} {J.\ Phys.\ A: Math.\ and General}\ }\textbf {\bibinfo {volume} {31}},\ \bibinfo {pages} {657--670} (\bibinfo {year} {1998})}\BibitemShut {NoStop}%
\bibitem [{\citenamefont {Child}, \citenamefont {Weston},\ and\ \citenamefont {Tennyson}(1999)}]{Child1999}%
  \BibitemOpen
  \bibfield  {author} {\bibinfo {author} {\bibfnamefont {M.~S.}\ \bibnamefont {Child}}, \bibinfo {author} {\bibfnamefont {T.}~\bibnamefont {Weston}}, \ and\ \bibinfo {author} {\bibfnamefont {J.}~\bibnamefont {Tennyson}},\ }\bibfield  {title} {\enquote {\bibinfo {title} {Quantum monodromy in the spectrum of {H$_2$O} and other systems: New insight into the level structure of quasi-linear molecules},}\ }\href {\doibase 10.1080/00268979909482971} {\bibfield  {journal} {\bibinfo  {journal} {Mol.\ Phys.}\ }\textbf {\bibinfo {volume} {96}},\ \bibinfo {pages} {371--379} (\bibinfo {year} {1999})}\BibitemShut {NoStop}%
\bibitem [{\citenamefont {Caprio}, \citenamefont {Cejnar},\ and\ \citenamefont {Iachello}(2008)}]{Caprio2008}%
  \BibitemOpen
  \bibfield  {author} {\bibinfo {author} {\bibfnamefont {M.~A.}\ \bibnamefont {Caprio}}, \bibinfo {author} {\bibfnamefont {P.}~\bibnamefont {Cejnar}}, \ and\ \bibinfo {author} {\bibfnamefont {F.}~\bibnamefont {Iachello}},\ }\bibfield  {title} {\enquote {\bibinfo {title} {Excited state quantum phase transitions in many-body systems},}\ }\href@noop {} {\bibfield  {journal} {\bibinfo  {journal} {Annals of Physics}\ }\textbf {\bibinfo {volume} {323}},\ \bibinfo {pages} {1106 -- 1135} (\bibinfo {year} {2008})}\BibitemShut {NoStop}%
\bibitem [{\citenamefont {Cejnar}\ and\ \citenamefont {Stransky}(2008)}]{pavels}%
  \BibitemOpen
  \bibfield  {author} {\bibinfo {author} {\bibfnamefont {P.}~\bibnamefont {Cejnar}}\ and\ \bibinfo {author} {\bibfnamefont {P.}~\bibnamefont {Stransky}},\ }\bibfield  {title} {\enquote {\bibinfo {title} {{Impact of quantum phase transitions on excited-level dynamics}},}\ }\href@noop {} {\bibfield  {journal} {\bibinfo  {journal} {Phys.\ Rev.\ E}\ }\textbf {\bibinfo {volume} {{78}}} (\bibinfo {year} {{2008}})}\BibitemShut {NoStop}%
\bibitem [{\citenamefont {Cejnar}\ \emph {et~al.}(2021)\citenamefont {Cejnar}, \citenamefont {Stránský}, \citenamefont {Macek},\ and\ \citenamefont {Kloc}}]{Cejnar2021}%
  \BibitemOpen
  \bibfield  {author} {\bibinfo {author} {\bibfnamefont {P.}~\bibnamefont {Cejnar}}, \bibinfo {author} {\bibfnamefont {P.}~\bibnamefont {Stránský}}, \bibinfo {author} {\bibfnamefont {M.}~\bibnamefont {Macek}}, \ and\ \bibinfo {author} {\bibfnamefont {M.}~\bibnamefont {Kloc}},\ }\bibfield  {title} {\enquote {\bibinfo {title} {Excited-state quantum phase transitions},}\ }\href {\doibase 10.1088/1751-8121/abdfe8} {\bibfield  {journal} {\bibinfo  {journal} {Journal of Physics A: Mathematical and Theoretical}\ }\textbf {\bibinfo {volume} {54}},\ \bibinfo {pages} {133001} (\bibinfo {year} {2021})}\BibitemShut {NoStop}%
\bibitem [{\citenamefont {Larese}\ and\ \citenamefont {Iachello}(2011)}]{larese2011}%
  \BibitemOpen
  \bibfield  {author} {\bibinfo {author} {\bibfnamefont {D.}~\bibnamefont {Larese}}\ and\ \bibinfo {author} {\bibfnamefont {F.}~\bibnamefont {Iachello}},\ }\bibfield  {title} {\enquote {\bibinfo {title} {A study of quantum phase transitions and quantum monodromy in the bending motion of non-rigid molecules},}\ }\href@noop {} {\bibfield  {journal} {\bibinfo  {journal} {J. Mol. Struct.}\ }\textbf {\bibinfo {volume} {1006}},\ \bibinfo {pages} {611--28} (\bibinfo {year} {2011})}\BibitemShut {NoStop}%
\bibitem [{\citenamefont {Larese}, \citenamefont {Pérez-Bernal},\ and\ \citenamefont {Iachello}(2013)}]{LARESE2013310}%
  \BibitemOpen
  \bibfield  {author} {\bibinfo {author} {\bibfnamefont {D.}~\bibnamefont {Larese}}, \bibinfo {author} {\bibfnamefont {F.}~\bibnamefont {Pérez-Bernal}}, \ and\ \bibinfo {author} {\bibfnamefont {F.}~\bibnamefont {Iachello}},\ }\bibfield  {title} {\enquote {\bibinfo {title} {Signatures of quantum phase transitions and excited state quantum phase transitions in the vibrational bending dynamics of triatomic molecules},}\ }\href {\doibase https://doi.org/10.1016/j.molstruc.2013.08.020} {\bibfield  {journal} {\bibinfo  {journal} {Journal of Molecular Structure}\ }\textbf {\bibinfo {volume} {1051}},\ \bibinfo {pages} {310--327} (\bibinfo {year} {2013})}\BibitemShut {NoStop}%
\bibitem [{\citenamefont {Khalouf-Rivera}, \citenamefont {P{\'e}rez-Bernal},\ and\ \citenamefont {Carvajal}(2021)}]{KHALOUF2021}%
  \BibitemOpen
  \bibfield  {author} {\bibinfo {author} {\bibfnamefont {J.}~\bibnamefont {Khalouf-Rivera}}, \bibinfo {author} {\bibfnamefont {F.}~\bibnamefont {P{\'e}rez-Bernal}}, \ and\ \bibinfo {author} {\bibfnamefont {M.}~\bibnamefont {Carvajal}},\ }\bibfield  {title} {\enquote {\bibinfo {title} {Excited state quantum phase transitions in the bending spectra of molecules},}\ }\href {\doibase https://doi.org/10.1016/j.jqsrt.2020.107436} {\bibfield  {journal} {\bibinfo  {journal} {Journal of Quantitative Spectroscopy and Radiative Transfer}\ }\textbf {\bibinfo {volume} {261}},\ \bibinfo {pages} {107436} (\bibinfo {year} {2021})}\BibitemShut {NoStop}%
\bibitem [{\citenamefont {Khalouf-Rivera}, \citenamefont {Carvajal},\ and\ \citenamefont {P\'erez-Bernal}(2022)}]{khalouf2022}%
  \BibitemOpen
  \bibfield  {author} {\bibinfo {author} {\bibfnamefont {J.}~\bibnamefont {Khalouf-Rivera}}, \bibinfo {author} {\bibfnamefont {M.}~\bibnamefont {Carvajal}}, \ and\ \bibinfo {author} {\bibfnamefont {F.}~\bibnamefont {P\'erez-Bernal}},\ }\bibfield  {title} {\enquote {\bibinfo {title} {Quantum fidelity susceptibility in excited state quantum phase transitions: Application to the bending spectra of nonrigid molecules},}\ }\href@noop {} {\bibfield  {journal} {\bibinfo  {journal} {SciPost Phys.}\ }\textbf {\bibinfo {volume} {12}},\ \bibinfo {pages} {002} (\bibinfo {year} {2022})}\BibitemShut {NoStop}%
\bibitem [{\citenamefont {Novotn\'y}\ and\ \citenamefont {Str\'ansk\'y}(2023)}]{Novotny2023}%
  \BibitemOpen
  \bibfield  {author} {\bibinfo {author} {\bibfnamefont {J.}~\bibnamefont {Novotn\'y}}\ and\ \bibinfo {author} {\bibfnamefont {P.}~\bibnamefont {Str\'ansk\'y}},\ }\bibfield  {title} {\enquote {\bibinfo {title} {Relative asymptotic oscillations of the out-of-time-ordered correlator as a quantum chaos indicator},}\ }\href {\doibase 10.1103/PhysRevE.107.054220} {\bibfield  {journal} {\bibinfo  {journal} {Phys. Rev. E}\ }\textbf {\bibinfo {volume} {107}},\ \bibinfo {pages} {054220} (\bibinfo {year} {2023})}\BibitemShut {NoStop}%
\bibitem [{\citenamefont {Sánchez-Castellanos}\ \emph {et~al.}(2012{\natexlab{a}})\citenamefont {Sánchez-Castellanos}, \citenamefont {Lemus}, \citenamefont {Carvajal},\ and\ \citenamefont {Pérez-Bernal}}]{Mariano2012}%
  \BibitemOpen
  \bibfield  {author} {\bibinfo {author} {\bibfnamefont {M.}~\bibnamefont {Sánchez-Castellanos}}, \bibinfo {author} {\bibfnamefont {R.}~\bibnamefont {Lemus}}, \bibinfo {author} {\bibfnamefont {M.}~\bibnamefont {Carvajal}}, \ and\ \bibinfo {author} {\bibfnamefont {F.}~\bibnamefont {Pérez-Bernal}},\ }\bibfield  {title} {\enquote {\bibinfo {title} {The potential energy surface of {CO$_2$} from an algebraic approach},}\ }\href {\doibase https://doi.org/10.1002/qua.24141} {\bibfield  {journal} {\bibinfo  {journal} {International Journal of Quantum Chemistry}\ }\textbf {\bibinfo {volume} {112}},\ \bibinfo {pages} {3498--3507} (\bibinfo {year} {2012}{\natexlab{a}})}\BibitemShut {NoStop}%
\bibitem [{\citenamefont {Sánchez-Castellanos}\ \emph {et~al.}(2012{\natexlab{b}})\citenamefont {Sánchez-Castellanos}, \citenamefont {Lemus}, \citenamefont {Carvajal}, \citenamefont {Pérez-Bernal},\ and\ \citenamefont {Fernández}}]{SanchezCastellanos2012}%
  \BibitemOpen
  \bibfield  {author} {\bibinfo {author} {\bibfnamefont {M.}~\bibnamefont {Sánchez-Castellanos}}, \bibinfo {author} {\bibfnamefont {R.}~\bibnamefont {Lemus}}, \bibinfo {author} {\bibfnamefont {M.}~\bibnamefont {Carvajal}}, \bibinfo {author} {\bibfnamefont {F.}~\bibnamefont {Pérez-Bernal}}, \ and\ \bibinfo {author} {\bibfnamefont {J.}~\bibnamefont {Fernández}},\ }\bibfield  {title} {\enquote {\bibinfo {title} {A study of the {Raman} spectrum of {CO$_2$} using an algebraic approach},}\ }\href@noop {} {\bibfield  {journal} {\bibinfo  {journal} {Chemical Physics Letters}\ }\textbf {\bibinfo {volume} {554}},\ \bibinfo {pages} {208--213} (\bibinfo {year} {2012}{\natexlab{b}})}\BibitemShut {NoStop}%
\bibitem [{\citenamefont {Estévez-Fregoso}\ and\ \citenamefont {Lemus}(2018)}]{Estevez-Fregoso2018}%
  \BibitemOpen
  \bibfield  {author} {\bibinfo {author} {\bibfnamefont {M.~M.}\ \bibnamefont {Estévez-Fregoso}}\ and\ \bibinfo {author} {\bibfnamefont {R.}~\bibnamefont {Lemus}},\ }\bibfield  {title} {\enquote {\bibinfo {title} {Connection between the su(3) algebraic and configuration spaces: bending modes of linear molecules},}\ }\href@noop {} {\bibfield  {journal} {\bibinfo  {journal} {Molecular Physics}\ }\textbf {\bibinfo {volume} {116}},\ \bibinfo {pages} {2374--2395} (\bibinfo {year} {2018})}\BibitemShut {NoStop}%
\bibitem [{\citenamefont {Bermúdez-Montaña}\ \emph {et~al.}(2022)\citenamefont {Bermúdez-Montaña}, \citenamefont {Rodríguez-Arcos}, \citenamefont {Carvajal}, \citenamefont {Ostertag-Henning},\ and\ \citenamefont {Lemus}}]{Marisol2022}%
  \BibitemOpen
  \bibfield  {author} {\bibinfo {author} {\bibfnamefont {M.}~\bibnamefont {Bermúdez-Montaña}}, \bibinfo {author} {\bibfnamefont {M.}~\bibnamefont {Rodríguez-Arcos}}, \bibinfo {author} {\bibfnamefont {M.}~\bibnamefont {Carvajal}}, \bibinfo {author} {\bibfnamefont {C.}~\bibnamefont {Ostertag-Henning}}, \ and\ \bibinfo {author} {\bibfnamefont {R.}~\bibnamefont {Lemus}},\ }\bibfield  {title} {\enquote {\bibinfo {title} {Algebraic vibrational description of the symmetric isotopologues of {CO$_2$}: {$^{13}$C$^{16}$O$_2$}, {$^{12}$C$^{18}$O$_2$} and {$^{12}$C$^{17}$O$_2$}},}\ }\href {\doibase https://doi.org/10.1016/j.chemphys.2022.111481} {\bibfield  {journal} {\bibinfo  {journal} {Chemical Physics}\ }\textbf {\bibinfo {volume} {557}},\ \bibinfo {pages} {111481} (\bibinfo {year} {2022})}\BibitemShut {NoStop}%
\bibitem [{\citenamefont {Bermúdez-Monta{\~n}a}\ \emph {et~al.}(2023)\citenamefont {Bermúdez-Monta{\~n}a}, \citenamefont {Rodríguez-Arcos}, \citenamefont {Carvajal}, \citenamefont {Ostertag-Henning},\ and\ \citenamefont {Lemus}}]{Marisol2023}%
  \BibitemOpen
  \bibfield  {author} {\bibinfo {author} {\bibfnamefont {M.}~\bibnamefont {Bermúdez-Monta{\~n}a}}, \bibinfo {author} {\bibfnamefont {M.}~\bibnamefont {Rodríguez-Arcos}}, \bibinfo {author} {\bibfnamefont {M.}~\bibnamefont {Carvajal}}, \bibinfo {author} {\bibfnamefont {C.}~\bibnamefont {Ostertag-Henning}}, \ and\ \bibinfo {author} {\bibfnamefont {R.}~\bibnamefont {Lemus}},\ }\bibfield  {title} {\enquote {\bibinfo {title} {A spectroscopic description of asymmetric isotopologues of {CO$_2$}},}\ }\href {\doibase 10.1021/acs.jpca.3c00890} {\bibfield  {journal} {\bibinfo  {journal} {The Journal of Physical Chemistry A}\ }\textbf {\bibinfo {volume} {127}},\ \bibinfo {pages} {6357--6376} (\bibinfo {year} {2023})}\BibitemShut {NoStop}%
\bibitem [{\citenamefont {{Suárez}}, \citenamefont {{Guzmán-Juárez}},\ and\ \citenamefont {Lemus}(2024)}]{Lemus2024}%
  \BibitemOpen
  \bibfield  {author} {\bibinfo {author} {\bibfnamefont {E.}~\bibnamefont {{Suárez}}}, \bibinfo {author} {\bibfnamefont {O.}~\bibnamefont {{Guzmán-Juárez}}}, \ and\ \bibinfo {author} {\bibfnamefont {R.}~\bibnamefont {Lemus}},\ }\bibfield  {title} {\enquote {\bibinfo {title} {A general local algebraic approach for molecules with normal mode behavior: Application to {FCN}},}\ }\href@noop {} {\bibfield  {journal} {\bibinfo  {journal} {Computational and Theoretical Chemistry}\ }\textbf {\bibinfo {volume} {1244}},\ \bibinfo {pages} {115069} (\bibinfo {year} {2024})}\BibitemShut {NoStop}%
\bibitem [{\citenamefont {Lemus}\ \emph {et~al.}(2014)\citenamefont {Lemus}, \citenamefont {S\'anchez-Castellanos}, \citenamefont {P\'erez-Bernal}, \citenamefont {Fern\'andez},\ and\ \citenamefont {Carvajal}}]{Lemus2014}%
  \BibitemOpen
  \bibfield  {author} {\bibinfo {author} {\bibfnamefont {R.}~\bibnamefont {Lemus}}, \bibinfo {author} {\bibfnamefont {M.}~\bibnamefont {S\'anchez-Castellanos}}, \bibinfo {author} {\bibfnamefont {F.}~\bibnamefont {P\'erez-Bernal}}, \bibinfo {author} {\bibfnamefont {J.~M.}\ \bibnamefont {Fern\'andez}}, \ and\ \bibinfo {author} {\bibfnamefont {M.}~\bibnamefont {Carvajal}},\ }\bibfield  {title} {\enquote {\bibinfo {title} {{Simulation of the {Raman} Spectra of {CO$_2$}: Bridging the Gap Between Algebraic Models and Experimental Spectra}},}\ }\href@noop {} {\bibfield  {journal} {\bibinfo  {journal} {J.\ Chem.\ Phys.}\ }\textbf {\bibinfo {volume} {141}},\ \bibinfo {pages} {054--306} (\bibinfo {year} {2014})}\BibitemShut {NoStop}%
\bibitem [{\citenamefont {Berm\'udez-Monta{\~n}a}\ \emph {et~al.}(2020)\citenamefont {Berm\'udez-Monta{\~n}a}, \citenamefont {Carvajal}, \citenamefont {P\'erez-Bernal},\ and\ \citenamefont {Lemus}}]{Marisol2020}%
  \BibitemOpen
  \bibfield  {author} {\bibinfo {author} {\bibfnamefont {M.}~\bibnamefont {Berm\'udez-Monta{\~n}a}}, \bibinfo {author} {\bibfnamefont {M.}~\bibnamefont {Carvajal}}, \bibinfo {author} {\bibfnamefont {F.}~\bibnamefont {P\'erez-Bernal}}, \ and\ \bibinfo {author} {\bibfnamefont {R.}~\bibnamefont {Lemus}},\ }\bibfield  {title} {\enquote {\bibinfo {title} {{An Algebraic Alternative for the Accurate Simulation of {CO$_2$} {Raman} Spectra}},}\ }\href@noop {} {\bibfield  {journal} {\bibinfo  {journal} {J.\ Raman Spectrosc.}\ }\textbf {\bibinfo {volume} {51}},\ \bibinfo {pages} {569--583} (\bibinfo {year} {2020})}\BibitemShut {NoStop}%
\bibitem [{\citenamefont {Santos}\ and\ \citenamefont {P\'erez-Bernal}(2015)}]{LSantos2015}%
  \BibitemOpen
  \bibfield  {author} {\bibinfo {author} {\bibfnamefont {L.~F.}\ \bibnamefont {Santos}}\ and\ \bibinfo {author} {\bibfnamefont {F.}~\bibnamefont {P\'erez-Bernal}},\ }\bibfield  {title} {\enquote {\bibinfo {title} {Structure of eigenstates and quench dynamics at an excited-state quantum phase transition},}\ }\href {\doibase 10.1103/PhysRevA.92.050101} {\bibfield  {journal} {\bibinfo  {journal} {Phys. Rev. A}\ }\textbf {\bibinfo {volume} {92}},\ \bibinfo {pages} {050101} (\bibinfo {year} {2015})}\BibitemShut {NoStop}%
\bibitem [{\citenamefont {Santos}, \citenamefont {T\'avora},\ and\ \citenamefont {P\'erez-Bernal}(2016)}]{Santos2016}%
  \BibitemOpen
  \bibfield  {author} {\bibinfo {author} {\bibfnamefont {L.~F.}\ \bibnamefont {Santos}}, \bibinfo {author} {\bibfnamefont {M.}~\bibnamefont {T\'avora}}, \ and\ \bibinfo {author} {\bibfnamefont {F.}~\bibnamefont {P\'erez-Bernal}},\ }\bibfield  {title} {\enquote {\bibinfo {title} {Excited-state quantum phase transitions in many-body systems with infinite-range interaction: Localization, dynamics, and bifurcation},}\ }\href {\doibase 10.1103/PhysRevA.94.012113} {\bibfield  {journal} {\bibinfo  {journal} {Phys. Rev. A}\ }\textbf {\bibinfo {volume} {94}},\ \bibinfo {pages} {012113} (\bibinfo {year} {2016})}\BibitemShut {NoStop}%
\bibitem [{\citenamefont {Khalouf-Rivera}, \citenamefont {P\'erez-Bernal},\ and\ \citenamefont {Carvajal}(2022)}]{jamil2022}%
  \BibitemOpen
  \bibfield  {author} {\bibinfo {author} {\bibfnamefont {J.}~\bibnamefont {Khalouf-Rivera}}, \bibinfo {author} {\bibfnamefont {F.}~\bibnamefont {P\'erez-Bernal}}, \ and\ \bibinfo {author} {\bibfnamefont {M.}~\bibnamefont {Carvajal}},\ }\bibfield  {title} {\enquote {\bibinfo {title} {Anharmonicity-induced excited-state quantum phase transition in the symmetric phase of the two-dimensional limit of the vibron model},}\ }\href {\doibase 10.1103/PhysRevA.105.032215} {\bibfield  {journal} {\bibinfo  {journal} {Phys. Rev. A}\ }\textbf {\bibinfo {volume} {105}},\ \bibinfo {pages} {032215} (\bibinfo {year} {2022})}\BibitemShut {NoStop}%
\bibitem [{\citenamefont {P\'erez-Bernal}\ and\ \citenamefont {\'Alvarez-Bajo}(2010)}]{AlvarezBajo2010}%
  \BibitemOpen
  \bibfield  {author} {\bibinfo {author} {\bibfnamefont {F.}~\bibnamefont {P\'erez-Bernal}}\ and\ \bibinfo {author} {\bibfnamefont {O.}~\bibnamefont {\'Alvarez-Bajo}},\ }\bibfield  {title} {\enquote {\bibinfo {title} {Anharmonicity effects in the bosonic u(2)-so(3) excited-state quantum phase transition},}\ }\href {\doibase 10.1103/PhysRevA.81.050101} {\bibfield  {journal} {\bibinfo  {journal} {Phys. Rev. A}\ }\textbf {\bibinfo {volume} {81}},\ \bibinfo {pages} {050101} (\bibinfo {year} {2010})}\BibitemShut {NoStop}%
\bibitem [{\citenamefont {Newville}\ \emph {et~al.}(2016)\citenamefont {Newville}, \citenamefont {Stensitzki}, \citenamefont {Allen}, \citenamefont {Rawlik}, \citenamefont {Ingargiola},\ and\ \citenamefont {Nelson}}]{newville2016lmfit}%
  \BibitemOpen
  \bibfield  {author} {\bibinfo {author} {\bibfnamefont {M.}~\bibnamefont {Newville}}, \bibinfo {author} {\bibfnamefont {T.}~\bibnamefont {Stensitzki}}, \bibinfo {author} {\bibfnamefont {D.~B.}\ \bibnamefont {Allen}}, \bibinfo {author} {\bibfnamefont {M.}~\bibnamefont {Rawlik}}, \bibinfo {author} {\bibfnamefont {A.}~\bibnamefont {Ingargiola}}, \ and\ \bibinfo {author} {\bibfnamefont {A.}~\bibnamefont {Nelson}},\ }\bibfield  {title} {\enquote {\bibinfo {title} {{LMFIT:} non-linear least-square minimization and curve-fitting for {Python}},}\ }\href@noop {} {\bibfield  {journal} {\bibinfo  {journal} {Astrophysics Source Code Library}\ ,\ \bibinfo {pages} {ascl--1606}} (\bibinfo {year} {2016})}\BibitemShut {NoStop}%
\bibitem [{\citenamefont {Zelevinsky}\ \emph {et~al.}(1996)\citenamefont {Zelevinsky}, \citenamefont {Brown}, \citenamefont {Frazier},\ and\ \citenamefont {Horoi}}]{ZELEVINSKY1996}%
  \BibitemOpen
  \bibfield  {author} {\bibinfo {author} {\bibfnamefont {V.}~\bibnamefont {Zelevinsky}}, \bibinfo {author} {\bibfnamefont {B.}~\bibnamefont {Brown}}, \bibinfo {author} {\bibfnamefont {N.}~\bibnamefont {Frazier}}, \ and\ \bibinfo {author} {\bibfnamefont {M.}~\bibnamefont {Horoi}},\ }\bibfield  {title} {\enquote {\bibinfo {title} {The nuclear shell model as a testing ground for many-body quantum chaos},}\ }\href {\doibase https://doi.org/10.1016/S0370-1573(96)00007-5} {\bibfield  {journal} {\bibinfo  {journal} {Physics Reports}\ }\textbf {\bibinfo {volume} {276}},\ \bibinfo {pages} {85--176} (\bibinfo {year} {1996})}\BibitemShut {NoStop}%
\end{thebibliography}%

\end{document}